\documentclass[aps,prl,twocolumn,superscriptaddress,floatfix]{revtex4-1}

\usepackage{dblfloatfix}

\usepackage{pifont}
\usepackage{subfigure}
\usepackage{amssymb}
\usepackage{amsmath}
\usepackage{times}
\usepackage{graphicx}
\usepackage{epsfig}
\usepackage{xcolor,dsfont}

\RequirePackage{color}

\definecolor{MyDarkGreen}{rgb}{0.02,0.60,0.06}

\definecolor{kellygreen}{rgb}{0.3, 0.73, 0.09}
\definecolor{deeppink}{rgb}{1.0, 0.08, 0.58}

\graphicspath{{figures}}

\graphicspath{{figures}}

\begin{document}

\newcommand{\ket}[1]{\ensuremath{|#1\rangle}}
\newcommand{\wket}[1]{\ensuremath{\color{white}{|}#1\color{white}{\rangle}}}
\newcommand{\bra}[1]{{\langle #1|}}
\def\avg#1{\left\langle#1\right\rangle}
\def\bra#1{\left\langle#1\right|}
\def\ket#1{\left|#1\right\rangle}
\def\Eq#1{Eq. {\eqref{#1}}}
\def\s{\sigma}
\def\t{\tau}
\def\d{\delta}
\def\be{\begin{equation}}
\def\ee{\end{equation}}
\def\o{\omega}
\def\bea{\begin{eqnarray}}
\def\eea{\end{eqnarray}}
\def\L{{\cal L} }

\title{\color{black}{Topological Quantum Many-Body Scars in Quantum Dimer Models on the Kagome Lattice}}

\author{Julia Wildeboer}
\affiliation{Department of Physics, Arizona State University, Tempe, Arizona 85287-1504, USA}
\author{Alexander Seidel}
\email[Corresponding author: ]{seidel@physics.wustl.edu}
\affiliation{Department of Physics, Washington University, St. Louis, Missouri 63130, USA}
\author{N. S. Srivatsa}
\affiliation{Max-Planck-Institute for the Physics of Complex Systems, D-01187 Dresden, Germany}
\author{Anne E. B. Nielsen}
\affiliation{Max-Planck-Institute for the Physics of Complex Systems, D-01187 Dresden, Germany}
\affiliation{Department of Physics and Astronomy, Aarhus University, DK-8000 Aarhus C, Denmark}
\author{Onur Erten}
\affiliation{Department of Physics, Arizona State University, Tempe, Arizona 85287-1504, USA}

\begin{abstract}
We present a class of quantum dimer models on the kagome lattice with 
full translational invariance that feature a quantum many-body scar state 
of analytically known entanglement properties within their spectra. 
Using exact diagonalization on lattices of up to 60 sites, we show that 
non-scar states conform to the eigenstate thermalization hypothesis. 
Specifically, we show that energies are distributed according 
to the Gaussian ensemble expected of their respective symmetry sector, 
illustrate the existence of the scar from bipartite entanglement properties,
  and demonstrate revival phenomena in studies of fidelity dynamics. 
\end{abstract}

\maketitle
Properties of strongly interacting quantum systems away from equilibrium 
are attracting a lot of attention in contemporary condensed matter theory. 
Progress in experiments \cite{bernien2017probing,kinoshita2006quantum,schreiber2015observation,smith2016many,kucsko} 
now allows for the preparation and study of quantum 
many-body systems that are well isolated from the environment, thereby giving access 
to non-equilibrium phenomena. 
One such phenomenon is given by the so-called quantum-many body scar states that were 
recently identified to be responsible for the unusual dynamics unexpectedly observed 
in one-dimensional Rydberg atom systems \cite{bernien2017probing,turner2018quantum,turner2018weak,lin2019exact}. 

Progress concerning theoretical studies poses an 
interesting and challenging task since the widely employed statistical mechanics tools 
fail to capture and describe relevant properties in out-of-equilibrium systems, e.g., the 
concept of the eigenstate thermalization hypothesis (ETH) breaks down. 
The ETH \cite{deutsch,berry,srednicki,rigol,deutsch_2018}  
postulates that generic closed quantum many-body systems exhibit ergodicity. 
Nowadays it is widely known that there are several important exceptions to this paradigm 
including but not limited to strong ergodicity breaking many-body localized states \cite{review_abanin,review_alet,theveniaut2019many}
and weak ergodicity breaking quantum many-body states \cite{moud,turner2018weak,turner2018quantum,moud2,choi,vedika,lin2019exact,ho,ok,lee2020exact,iadecola2019quantum},
where only a finite number of eigenstates, the scar states, 
break ergodicity while the majority of states respects the ETH. 

In this Letter, we focus on the latter case. 
Multiple possible scenarios are being investigated in the current literature.  
Progress has predominantly been made in one-dimensional systems such as the 
celebrated PXP-model \cite{turner2018weak,turner2018quantum,choi,vedika,lin2019exact,ho,iadecola2019quantum,
shiraishi2019connection,michailidis2019slow,moudgalya2019quantum}
realized in the Rydberg atoms experiment \cite{bernien2017probing}. 
Further advances were made by analytically constructing scar eigenstates \cite{moud,moud2} 
in Affleck-Kennedy-Lieb-Tasaki (AKLT) spin chains \cite{aklt} and in the fractional quantum Hall 
thin-torus limit \cite{moudgalya2019quantum}. 
Recently a few 2D systems have come under investigation \cite{ok,lee2020exact,lin2020quantum}. 
The literature currently offers several possible scenarios with respect to the mechanism giving rise to the quantum scars phenomenon, 
ranging from proximity to integrability \cite{vedika}, ``embedded'' SU(2) dynamics \cite{choi,mori} 
and magnon condensation \cite{iadecola2019quantum}. 
At present, there seems to be a scarcity of models on two-dimensional lattices with translational invariance and 
isolated quantum many-body scars that are numerically well-documented in terms of level statistics, entanglement 
entropy, and equilibration dynamics. 
Indeed, numerical studies are often limited by the size of the configuration space involved, 
particularly so in higher dimensions. 
In the present work, we examine a simple strategy to introduce an analytically known scar state given any class of frustration free 
Hamiltonians, of which there are many examples in the literature. Given this, we focus on quantum dimer models 
for their relatively moderate (though still exponential) scaling between system size and Hilbert space dimension.
Though generalization is straightforward, we will focus on the kagome lattice, which unites several advantages in this context:
Favorable Hilbert space size scaling ($2^{\sf (lattice \;sites)/3}$), analytically accessible entanglement properties of the scar state,
and a large number of natural parameters per unit cell \footnote{
The present strategy can be applied to other dimer models on other lattices, where lack
of multiple free parameters per unit cell may be overcome by doubling the unit cell.
However, to the best of our knowledge, analytic control over {\em entanglement} properties is 
unique to the kagome case.}. We also note that dimer-related models in the kagome geometry
have recently been argued to offer an attractive route to the experimental realization of exotic physics \cite{verresen}.

Our main results are as follows: $(i)$ Following a general strategy, we construct a class of 
quantum dimer models on the kagome lattice containing quantum many-body scar states in 
their spectrum that provably violate the ETH, having sub-volume entanglement. 
$(ii)$ Making use of the favorable Hilbert-space scaling of kagome dimer models, 
we numerically demonstrate that the remaining states in the spectrum thermalize 
by analyzing their level statistics and entanglement entropy.  We further study fidelity dynamics,
demonstrating the presence of scar states in the spectrum and their effects on thermalization.

{\it Quantum dimer models. ---} 
Rokhsar and Kivelson introduced  quantum dimer models (QDMs) \cite{rk} 
for the sake of capturing the essential topological features of the short-ranged variety of Anderson's  
resonating valence bond states in a model that is tractable. 
Originally designed to advance the understanding of high-temperature superconductors, 
quantum dimer models have played an increasing role in describing 
new and unusual emergent phenomena in many-body systems \cite{ms,msf,misguich,moessner2011introduction,wildeboer2020}. 
This includes, in particular, studies on many-body localization in constrained systems \cite{theveniaut2019many}. 
We now proceed by summarizing some key features of the quantum dimer model on the kagome lattice 
introduced by Misguich  {\it et al.} \cite{misguich}, 
before introducing a variant of this model that displays quantum many-body scars in its spectrum. 

The QDM is defined on a Hilbert space of distinct orthonormal states that represent the 
allowed hard-core dimer coverings of the lattice 
 such that  each site participates in exactly one dimer between nearest neighbors. 
The Hamiltonian is then defined by local matrix elements between dimer states, where we distinguish 
 ``potential'' terms, $V$, that are diagonal in the dimer basis and associate an 
interaction energy with various local arrangements of dimers, and ``kinetic'' terms, $t$, 
that facilitate a local rearrangement of a small number of dimers. This Letter solely focuses on the kagome lattice where all 
local interactions take place within twelve-site star-shaped cells, Table \ref{table3}. 

Graphically, the Hamiltonian is presented as:  
\begin{widetext}
\begin{eqnarray}\label{ham1}
{\cal H} = 
\sum_{\textrm{\ding{65}}}
\Bigg[\!&&-t_{1}\left( \ket{%
%
\setlength{\unitlength}{3947sp}
\begin{picture}(320,230)(-10,160)


 \color{magenta}
 \put (110,274){\circle*{1}}
 \put (120,274){\circle*{1}}
 \put (130,274){\circle*{1}}
 \put (140,274){\circle*{1}}
 \put (150,274){\circle*{1}}
 \put (160,274){\circle*{1}}
 \put (170,274){\circle*{1}}
 \put (180,274){\circle*{1}}
 \put (190,274){\circle*{1}}
 \put (195,274){\circle*{1}}

 \put (100,100){\circle*{1}}
 \put ( 95,110){\circle*{1}}
 \put ( 85,120){\circle*{1}}
 \put ( 80,130){\circle*{1}}
 \put ( 75,140){\circle*{1}}
 \put ( 70,150){\circle*{1}}
 \put ( 65,160){\circle*{1}}
 \put ( 60,170){\circle*{1}}
 \put ( 55,180){\circle*{1}}
 \put ( 50,187){\circle*{1}}

 \put (200,100){\circle*{1}}
 \put (205,110){\circle*{1}}
 \put (215,120){\circle*{1}}
 \put (220,130){\circle*{1}}
 \put (225,140){\circle*{1}}
 \put (230,150){\circle*{1}}
 \put (235,160){\circle*{1}}
 \put (240,170){\circle*{1}}
 \put (245,180){\circle*{1}}
 \put (250,187){\circle*{1}}

 \color{magenta}
 \put (150,13){\circle*{25}}
 \put (  0,100){\circle*{25}}
 \put (100,100){\circle*{25}}
 \put (200,100){\circle*{25}}
 \put (300,100){\circle*{25}}
 \put ( 50,187){\circle*{25}}
 \put (250,187){\circle*{25}}
 \put (  0,274){\circle*{25}}
 \put (100,274){\circle*{25}}
 \put (200,274){\circle*{25}}
 \put (300,274){\circle*{25}}
 \put (150,361){\circle*{25}}

 \color{black}
 \put (  0,274){\circle*{25}}
 \put (300,274){\circle*{25}}
 \put (  0,100){\circle*{25}}
 \put (300,100){\circle*{25}}
 \put (150, 13){\circle*{25}}
 \put (150,361){\circle*{25}}
\end{picture}}\bra{%
%
\setlength{\unitlength}{3947sp}
\begin{picture}(320,230)(-10,160)


 \color{magenta}
 \put (110,100){\circle*{1}}
 \put (120,100){\circle*{1}}
 \put (130,100){\circle*{1}}
 \put (140,100){\circle*{1}}
 \put (150,100){\circle*{1}}
 \put (160,100){\circle*{1}}
 \put (170,100){\circle*{1}}
 \put (180,100){\circle*{1}}
 \put (190,100){\circle*{1}}
 \put (195,100){\circle*{1}}

 \put ( 50,187){\circle*{1}}
 \put ( 55,197){\circle*{1}}
 \put ( 60,207){\circle*{1}}
 \put ( 65,217){\circle*{1}}
 \put ( 70,227){\circle*{1}}
 \put ( 75,237){\circle*{1}}
 \put ( 80,247){\circle*{1}}
 \put ( 85,257){\circle*{1}}
 \put ( 90,267){\circle*{1}}
 \put ( 95,274){\circle*{1}}

 \put (250,187){\circle*{1}}
 \put (245,197){\circle*{1}}
 \put (240,207){\circle*{1}}
 \put (235,217){\circle*{1}}
 \put (230,227){\circle*{1}}
 \put (225,237){\circle*{1}}
 \put (220,247){\circle*{1}}
 \put (215,257){\circle*{1}}
 \put (210,267){\circle*{1}}
 \put (205,274){\circle*{1}}

 \color{magenta}
 \put (150,13){\circle*{25}}
 \put (  0,100){\circle*{25}}
 \put (100,100){\circle*{25}}
 \put (200,100){\circle*{25}}
 \put (300,100){\circle*{25}}
 \put ( 50,187){\circle*{25}}
 \put (250,187){\circle*{25}}
 \put (  0,274){\circle*{25}}
 \put (100,274){\circle*{25}}
 \put (200,274){\circle*{25}}
 \put (300,274){\circle*{25}}
 \put (150,361){\circle*{25}}

 \color{black}
 \put (  0,274){\circle*{25}}
 \put (300,274){\circle*{25}}
 \put (  0,100){\circle*{25}}
 \put (300,100){\circle*{25}}
 \put (150, 13){\circle*{25}}
 \put (150,361){\circle*{25}} 
\end{picture}} + \ket{%
%
\setlength{\unitlength}{3947sp}
\begin{picture}(320,230)(-10,160)


 \color{magenta}
 \put (110,100){\circle*{1}}
 \put (120,100){\circle*{1}}
 \put (130,100){\circle*{1}}
 \put (140,100){\circle*{1}}
 \put (150,100){\circle*{1}}
 \put (160,100){\circle*{1}}
 \put (170,100){\circle*{1}}
 \put (180,100){\circle*{1}}
 \put (190,100){\circle*{1}}
 \put (195,100){\circle*{1}}

 \put ( 50,187){\circle*{1}}
 \put ( 55,197){\circle*{1}}
 \put ( 60,207){\circle*{1}}
 \put ( 65,217){\circle*{1}}
 \put ( 70,227){\circle*{1}}
 \put ( 75,237){\circle*{1}}
 \put ( 80,247){\circle*{1}}
 \put ( 85,257){\circle*{1}}
 \put ( 90,267){\circle*{1}}
 \put ( 95,274){\circle*{1}}

 \put (250,187){\circle*{1}}
 \put (245,197){\circle*{1}}
 \put (240,207){\circle*{1}}
 \put (235,217){\circle*{1}}
 \put (230,227){\circle*{1}}
 \put (225,237){\circle*{1}}
 \put (220,247){\circle*{1}}
 \put (215,257){\circle*{1}}
 \put (210,267){\circle*{1}}
 \put (205,274){\circle*{1}}

 \color{magenta}
 \put (150,13){\circle*{25}}
 \put (  0,100){\circle*{25}}
 \put (100,100){\circle*{25}}
 \put (200,100){\circle*{25}}
 \put (300,100){\circle*{25}}
 \put ( 50,187){\circle*{25}}
 \put (250,187){\circle*{25}}
 \put (  0,274){\circle*{25}}
 \put (100,274){\circle*{25}}
 \put (200,274){\circle*{25}}
 \put (300,274){\circle*{25}}
 \put (150,361){\circle*{25}}

 \color{black}
 \put (  0,274){\circle*{25}}
 \put (300,274){\circle*{25}}
 \put (  0,100){\circle*{25}}
 \put (300,100){\circle*{25}}
 \put (150, 13){\circle*{25}}
 \put (150,361){\circle*{25}} 
\end{picture}}\bra{%
%
\setlength{\unitlength}{3947sp}
\begin{picture}(320,230)(-10,160)


 \color{magenta}
 \put (110,274){\circle*{1}}
 \put (120,274){\circle*{1}}
 \put (130,274){\circle*{1}}
 \put (140,274){\circle*{1}}
 \put (150,274){\circle*{1}}
 \put (160,274){\circle*{1}}
 \put (170,274){\circle*{1}}
 \put (180,274){\circle*{1}}
 \put (190,274){\circle*{1}}
 \put (195,274){\circle*{1}}

 \put (100,100){\circle*{1}}
 \put ( 95,110){\circle*{1}}
 \put ( 85,120){\circle*{1}}
 \put ( 80,130){\circle*{1}}
 \put ( 75,140){\circle*{1}}
 \put ( 70,150){\circle*{1}}
 \put ( 65,160){\circle*{1}}
 \put ( 60,170){\circle*{1}}
 \put ( 55,180){\circle*{1}}
 \put ( 50,187){\circle*{1}}

 \put (200,100){\circle*{1}}
 \put (205,110){\circle*{1}}
 \put (215,120){\circle*{1}}
 \put (220,130){\circle*{1}}
 \put (225,140){\circle*{1}}
 \put (230,150){\circle*{1}}
 \put (235,160){\circle*{1}}
 \put (240,170){\circle*{1}}
 \put (245,180){\circle*{1}}
 \put (250,187){\circle*{1}}

 \color{magenta}
 \put (150,13){\circle*{25}}
 \put (  0,100){\circle*{25}}
 \put (100,100){\circle*{25}}
 \put (200,100){\circle*{25}}
 \put (300,100){\circle*{25}}
 \put ( 50,187){\circle*{25}}
 \put (250,187){\circle*{25}}
 \put (  0,274){\circle*{25}}
 \put (100,274){\circle*{25}}
 \put (200,274){\circle*{25}}
 \put (300,274){\circle*{25}}
 \put (150,361){\circle*{25}}

 \color{black}
 \put (  0,274){\circle*{25}}
 \put (300,274){\circle*{25}}
 \put (  0,100){\circle*{25}}
 \put (300,100){\circle*{25}}
 \put (150, 13){\circle*{25}}
 \put (150,361){\circle*{25}}
\end{picture}}\right)
+V_{1}\left(\ket{%
%
\setlength{\unitlength}{3947sp}
\begin{picture}(320,230)(-10,160)


 \color{magenta}
 \put (110,274){\circle*{1}}
 \put (120,274){\circle*{1}}
 \put (130,274){\circle*{1}}
 \put (140,274){\circle*{1}}
 \put (150,274){\circle*{1}}
 \put (160,274){\circle*{1}}
 \put (170,274){\circle*{1}}
 \put (180,274){\circle*{1}}
 \put (190,274){\circle*{1}}
 \put (195,274){\circle*{1}}

 \put (100,100){\circle*{1}}
 \put ( 95,110){\circle*{1}}
 \put ( 85,120){\circle*{1}}
 \put ( 80,130){\circle*{1}}
 \put ( 75,140){\circle*{1}}
 \put ( 70,150){\circle*{1}}
 \put ( 65,160){\circle*{1}}
 \put ( 60,170){\circle*{1}}
 \put ( 55,180){\circle*{1}}
 \put ( 50,187){\circle*{1}}

 \put (200,100){\circle*{1}}
 \put (205,110){\circle*{1}}
 \put (215,120){\circle*{1}}
 \put (220,130){\circle*{1}}
 \put (225,140){\circle*{1}}
 \put (230,150){\circle*{1}}
 \put (235,160){\circle*{1}}
 \put (240,170){\circle*{1}}
 \put (245,180){\circle*{1}}
 \put (250,187){\circle*{1}}

 \color{magenta}
 \put (150,13){\circle*{25}}
 \put (  0,100){\circle*{25}}
 \put (100,100){\circle*{25}}
 \put (200,100){\circle*{25}}
 \put (300,100){\circle*{25}}
 \put ( 50,187){\circle*{25}}
 \put (250,187){\circle*{25}}
 \put (  0,274){\circle*{25}}
 \put (100,274){\circle*{25}}
 \put (200,274){\circle*{25}}
 \put (300,274){\circle*{25}}
 \put (150,361){\circle*{25}}

 \color{black}
 \put (  0,274){\circle*{25}}
 \put (300,274){\circle*{25}}
 \put (  0,100){\circle*{25}}
 \put (300,100){\circle*{25}}
 \put (150, 13){\circle*{25}}
 \put (150,361){\circle*{25}}
\end{picture}}\bra{%
%
\setlength{\unitlength}{3947sp}
\begin{picture}(320,230)(-10,160)


 \color{magenta}
 \put (110,274){\circle*{1}}
 \put (120,274){\circle*{1}}
 \put (130,274){\circle*{1}}
 \put (140,274){\circle*{1}}
 \put (150,274){\circle*{1}}
 \put (160,274){\circle*{1}}
 \put (170,274){\circle*{1}}
 \put (180,274){\circle*{1}}
 \put (190,274){\circle*{1}}
 \put (195,274){\circle*{1}}

 \put (100,100){\circle*{1}}
 \put ( 95,110){\circle*{1}}
 \put ( 85,120){\circle*{1}}
 \put ( 80,130){\circle*{1}}
 \put ( 75,140){\circle*{1}}
 \put ( 70,150){\circle*{1}}
 \put ( 65,160){\circle*{1}}
 \put ( 60,170){\circle*{1}}
 \put ( 55,180){\circle*{1}}
 \put ( 50,187){\circle*{1}}

 \put (200,100){\circle*{1}}
 \put (205,110){\circle*{1}}
 \put (215,120){\circle*{1}}
 \put (220,130){\circle*{1}}
 \put (225,140){\circle*{1}}
 \put (230,150){\circle*{1}}
 \put (235,160){\circle*{1}}
 \put (240,170){\circle*{1}}
 \put (245,180){\circle*{1}}
 \put (250,187){\circle*{1}}

 \color{magenta}
 \put (150,13){\circle*{25}}
 \put (  0,100){\circle*{25}}
 \put (100,100){\circle*{25}}
 \put (200,100){\circle*{25}}
 \put (300,100){\circle*{25}}
 \put ( 50,187){\circle*{25}}
 \put (250,187){\circle*{25}}
 \put (  0,274){\circle*{25}}
 \put (100,274){\circle*{25}}
 \put (200,274){\circle*{25}}
 \put (300,274){\circle*{25}}
 \put (150,361){\circle*{25}}

 \color{black}
 \put (  0,274){\circle*{25}}
 \put (300,274){\circle*{25}}
 \put (  0,100){\circle*{25}}
 \put (300,100){\circle*{25}}
 \put (150, 13){\circle*{25}}
 \put (150,361){\circle*{25}}
\end{picture}} + \ket{%
%
\setlength{\unitlength}{3947sp}
\begin{picture}(320,230)(-10,160)


 \color{magenta}
 \put (110,100){\circle*{1}}
 \put (120,100){\circle*{1}}
 \put (130,100){\circle*{1}}
 \put (140,100){\circle*{1}}
 \put (150,100){\circle*{1}}
 \put (160,100){\circle*{1}}
 \put (170,100){\circle*{1}}
 \put (180,100){\circle*{1}}
 \put (190,100){\circle*{1}}
 \put (195,100){\circle*{1}}

 \put ( 50,187){\circle*{1}}
 \put ( 55,197){\circle*{1}}
 \put ( 60,207){\circle*{1}}
 \put ( 65,217){\circle*{1}}
 \put ( 70,227){\circle*{1}}
 \put ( 75,237){\circle*{1}}
 \put ( 80,247){\circle*{1}}
 \put ( 85,257){\circle*{1}}
 \put ( 90,267){\circle*{1}}
 \put ( 95,274){\circle*{1}}

 \put (250,187){\circle*{1}}
 \put (245,197){\circle*{1}}
 \put (240,207){\circle*{1}}
 \put (235,217){\circle*{1}}
 \put (230,227){\circle*{1}}
 \put (225,237){\circle*{1}}
 \put (220,247){\circle*{1}}
 \put (215,257){\circle*{1}}
 \put (210,267){\circle*{1}}
 \put (205,274){\circle*{1}}

 \color{magenta}
 \put (150,13){\circle*{25}}
 \put (  0,100){\circle*{25}}
 \put (100,100){\circle*{25}}
 \put (200,100){\circle*{25}}
 \put (300,100){\circle*{25}}
 \put ( 50,187){\circle*{25}}
 \put (250,187){\circle*{25}}
 \put (  0,274){\circle*{25}}
 \put (100,274){\circle*{25}}
 \put (200,274){\circle*{25}}
 \put (300,274){\circle*{25}}
 \put (150,361){\circle*{25}}

 \color{black}
 \put (  0,274){\circle*{25}}
 \put (300,274){\circle*{25}}
 \put (  0,100){\circle*{25}}
 \put (300,100){\circle*{25}}
 \put (150, 13){\circle*{25}}
 \put (150,361){\circle*{25}} 
\end{picture}}\bra{%
%
\setlength{\unitlength}{3947sp}
\begin{picture}(320,230)(-10,160)


 \color{magenta}
 \put (110,100){\circle*{1}}
 \put (120,100){\circle*{1}}
 \put (130,100){\circle*{1}}
 \put (140,100){\circle*{1}}
 \put (150,100){\circle*{1}}
 \put (160,100){\circle*{1}}
 \put (170,100){\circle*{1}}
 \put (180,100){\circle*{1}}
 \put (190,100){\circle*{1}}
 \put (195,100){\circle*{1}}

 \put ( 50,187){\circle*{1}}
 \put ( 55,197){\circle*{1}}
 \put ( 60,207){\circle*{1}}
 \put ( 65,217){\circle*{1}}
 \put ( 70,227){\circle*{1}}
 \put ( 75,237){\circle*{1}}
 \put ( 80,247){\circle*{1}}
 \put ( 85,257){\circle*{1}}
 \put ( 90,267){\circle*{1}}
 \put ( 95,274){\circle*{1}}

 \put (250,187){\circle*{1}}
 \put (245,197){\circle*{1}}
 \put (240,207){\circle*{1}}
 \put (235,217){\circle*{1}}
 \put (230,227){\circle*{1}}
 \put (225,237){\circle*{1}}
 \put (220,247){\circle*{1}}
 \put (215,257){\circle*{1}}
 \put (210,267){\circle*{1}}
 \put (205,274){\circle*{1}}

 \color{magenta}
 \put (150,13){\circle*{25}}
 \put (  0,100){\circle*{25}}
 \put (100,100){\circle*{25}}
 \put (200,100){\circle*{25}}
 \put (300,100){\circle*{25}}
 \put ( 50,187){\circle*{25}}
 \put (250,187){\circle*{25}}
 \put (  0,274){\circle*{25}}
 \put (100,274){\circle*{25}}
 \put (200,274){\circle*{25}}
 \put (300,274){\circle*{25}}
 \put (150,361){\circle*{25}}

 \color{black}
 \put (  0,274){\circle*{25}}
 \put (300,274){\circle*{25}}
 \put (  0,100){\circle*{25}}
 \put (300,100){\circle*{25}}
 \put (150, 13){\circle*{25}}
 \put (150,361){\circle*{25}} 
\end{picture}}\right)\nonumber
\\
&&+\sum_{r_{\textrm{rot}}}\Bigg[\! -t_{2} \left( \ket{%
%
\setlength{\unitlength}{3947sp}
\begin{picture}(320,230)(-10,160)


 \color{magenta}
 \put (10,100){\circle*{1}}
 \put (20,100){\circle*{1}}
 \put (30,100){\circle*{1}}
 \put (40,100){\circle*{1}}
 \put (50,100){\circle*{1}}
 \put (60,100){\circle*{1}}
 \put (70,100){\circle*{1}}
 \put (80,100){\circle*{1}}
 \put (90,100){\circle*{1}}
 \put (95,100){\circle*{1}}

 \put (210,274){\circle*{1}}
 \put (220,274){\circle*{1}}
 \put (230,274){\circle*{1}}
 \put (240,274){\circle*{1}}
 \put (250,274){\circle*{1}}
 \put (260,274){\circle*{1}}
 \put (270,274){\circle*{1}}
 \put (280,274){\circle*{1}}
 \put (290,274){\circle*{1}}
 \put (295,274){\circle*{1}}

 \put ( 50,187){\circle*{1}}
 \put ( 55,197){\circle*{1}}
 \put ( 60,207){\circle*{1}}
 \put ( 65,217){\circle*{1}}
 \put ( 70,227){\circle*{1}}
 \put ( 75,237){\circle*{1}}
 \put ( 80,247){\circle*{1}}
 \put ( 85,257){\circle*{1}}
 \put ( 90,267){\circle*{1}}
 \put ( 95,274){\circle*{1}}

 \put (200,100){\circle*{1}}
 \put (205,110){\circle*{1}}
 \put (210,120){\circle*{1}}
 \put (215,130){\circle*{1}}
 \put (220,140){\circle*{1}}
 \put (225,150){\circle*{1}}
 \put (230,160){\circle*{1}}
 \put (235,170){\circle*{1}}
 \put (240,180){\circle*{1}}
 \put (245,190){\circle*{1}}

 \color{magenta}
 \put (150,13){\circle*{25}}
 \put (  0,100){\circle*{25}}
 \put (100,100){\circle*{25}}
 \put (200,100){\circle*{25}}
 \put (300,100){\circle*{25}}
 \put ( 50,187){\circle*{25}}
 \put (250,187){\circle*{25}}
 \put (  0,274){\circle*{25}}
 \put (100,274){\circle*{25}}
 \put (200,274){\circle*{25}}
 \put (300,274){\circle*{25}}
 \put (150,361){\circle*{25}}

 \color{black}
 \put (150,13){\circle*{25}}
 \put (150,361){\circle*{25}}
 \put (  0,274){\circle*{25}}
 \put (300,100){\circle*{25}}
\end{picture}}\bra{%
%
\setlength{\unitlength}{3947sp}
\begin{picture}(320,230)(-10,160)


 \color{magenta}
 \put (110,100){\circle*{1}}
 \put (120,100){\circle*{1}}
 \put (130,100){\circle*{1}}
 \put (140,100){\circle*{1}}
 \put (150,100){\circle*{1}}
 \put (160,100){\circle*{1}}
 \put (170,100){\circle*{1}}
 \put (180,100){\circle*{1}}
 \put (190,100){\circle*{1}}
 \put (195,100){\circle*{1}}

 \put (110,274){\circle*{1}}
 \put (120,274){\circle*{1}}
 \put (130,274){\circle*{1}}
 \put (140,274){\circle*{1}}
 \put (150,274){\circle*{1}}
 \put (160,274){\circle*{1}}
 \put (170,274){\circle*{1}}
 \put (180,274){\circle*{1}}
 \put (190,274){\circle*{1}}
 \put (195,274){\circle*{1}}

 \put (  0,100){\circle*{1}}
 \put (  5,110){\circle*{1}}
 \put ( 10,120){\circle*{1}}
 \put ( 15,130){\circle*{1}}
 \put ( 20,140){\circle*{1}}
 \put ( 25,150){\circle*{1}}
 \put ( 30,160){\circle*{1}}
 \put ( 35,170){\circle*{1}}
 \put ( 40,180){\circle*{1}}
 \put ( 45,187){\circle*{1}}

 \put (250,187){\circle*{1}}
 \put (255,197){\circle*{1}}
 \put (260,207){\circle*{1}}
 \put (265,217){\circle*{1}}
 \put (270,227){\circle*{1}}
 \put (275,237){\circle*{1}}
 \put (280,247){\circle*{1}}
 \put (285,257){\circle*{1}}
 \put (290,267){\circle*{1}}
 \put (295,277){\circle*{1}}

 \color{magenta}
 \put (150,13){\circle*{25}}
 \put (  0,100){\circle*{25}}
 \put (100,100){\circle*{25}}
 \put (200,100){\circle*{25}}
 \put (300,100){\circle*{25}}
 \put ( 50,187){\circle*{25}}
 \put (250,187){\circle*{25}}
 \put (  0,274){\circle*{25}}
 \put (100,274){\circle*{25}}
 \put (200,274){\circle*{25}}
 \put (300,274){\circle*{25}}
 \put (150,361){\circle*{25}}

 \color{black}
 \put (150,13){\circle*{25}}
 \put (150,361){\circle*{25}}
 \put (  0,274){\circle*{25}}
 \put (300,100){\circle*{25}}
\end{picture}} + \ket{%
%
\setlength{\unitlength}{3947sp}
\begin{picture}(320,230)(-10,160)


 \color{magenta}
 \put (110,100){\circle*{1}}
 \put (120,100){\circle*{1}}
 \put (130,100){\circle*{1}}
 \put (140,100){\circle*{1}}
 \put (150,100){\circle*{1}}
 \put (160,100){\circle*{1}}
 \put (170,100){\circle*{1}}
 \put (180,100){\circle*{1}}
 \put (190,100){\circle*{1}}
 \put (195,100){\circle*{1}}

 \put (110,274){\circle*{1}}
 \put (120,274){\circle*{1}}
 \put (130,274){\circle*{1}}
 \put (140,274){\circle*{1}}
 \put (150,274){\circle*{1}}
 \put (160,274){\circle*{1}}
 \put (170,274){\circle*{1}}
 \put (180,274){\circle*{1}}
 \put (190,274){\circle*{1}}
 \put (195,274){\circle*{1}}

 \put (  0,100){\circle*{1}}
 \put (  5,110){\circle*{1}}
 \put ( 10,120){\circle*{1}}
 \put ( 15,130){\circle*{1}}
 \put ( 20,140){\circle*{1}}
 \put ( 25,150){\circle*{1}}
 \put ( 30,160){\circle*{1}}
 \put ( 35,170){\circle*{1}}
 \put ( 40,180){\circle*{1}}
 \put ( 45,187){\circle*{1}}

 \put (250,187){\circle*{1}}
 \put (255,197){\circle*{1}}
 \put (260,207){\circle*{1}}
 \put (265,217){\circle*{1}}
 \put (270,227){\circle*{1}}
 \put (275,237){\circle*{1}}
 \put (280,247){\circle*{1}}
 \put (285,257){\circle*{1}}
 \put (290,267){\circle*{1}}
 \put (295,277){\circle*{1}}

 \color{magenta}
 \put (150,13){\circle*{25}}
 \put (  0,100){\circle*{25}}
 \put (100,100){\circle*{25}}
 \put (200,100){\circle*{25}}
 \put (300,100){\circle*{25}}
 \put ( 50,187){\circle*{25}}
 \put (250,187){\circle*{25}}
 \put (  0,274){\circle*{25}}
 \put (100,274){\circle*{25}}
 \put (200,274){\circle*{25}}
 \put (300,274){\circle*{25}}
 \put (150,361){\circle*{25}}

 \color{black}
 \put (150,13){\circle*{25}}
 \put (150,361){\circle*{25}}
 \put (  0,274){\circle*{25}}
 \put (300,100){\circle*{25}}
\end{picture}}\bra{%
%
\setlength{\unitlength}{3947sp}
\begin{picture}(320,230)(-10,160)


 \color{magenta}
 \put (10,100){\circle*{1}}
 \put (20,100){\circle*{1}}
 \put (30,100){\circle*{1}}
 \put (40,100){\circle*{1}}
 \put (50,100){\circle*{1}}
 \put (60,100){\circle*{1}}
 \put (70,100){\circle*{1}}
 \put (80,100){\circle*{1}}
 \put (90,100){\circle*{1}}
 \put (95,100){\circle*{1}}

 \put (210,274){\circle*{1}}
 \put (220,274){\circle*{1}}
 \put (230,274){\circle*{1}}
 \put (240,274){\circle*{1}}
 \put (250,274){\circle*{1}}
 \put (260,274){\circle*{1}}
 \put (270,274){\circle*{1}}
 \put (280,274){\circle*{1}}
 \put (290,274){\circle*{1}}
 \put (295,274){\circle*{1}}

 \put ( 50,187){\circle*{1}}
 \put ( 55,197){\circle*{1}}
 \put ( 60,207){\circle*{1}}
 \put ( 65,217){\circle*{1}}
 \put ( 70,227){\circle*{1}}
 \put ( 75,237){\circle*{1}}
 \put ( 80,247){\circle*{1}}
 \put ( 85,257){\circle*{1}}
 \put ( 90,267){\circle*{1}}
 \put ( 95,274){\circle*{1}}

 \put (200,100){\circle*{1}}
 \put (205,110){\circle*{1}}
 \put (210,120){\circle*{1}}
 \put (215,130){\circle*{1}}
 \put (220,140){\circle*{1}}
 \put (225,150){\circle*{1}}
 \put (230,160){\circle*{1}}
 \put (235,170){\circle*{1}}
 \put (240,180){\circle*{1}}
 \put (245,190){\circle*{1}}

 \color{magenta}
 \put (150,13){\circle*{25}}
 \put (  0,100){\circle*{25}}
 \put (100,100){\circle*{25}}
 \put (200,100){\circle*{25}}
 \put (300,100){\circle*{25}}
 \put ( 50,187){\circle*{25}}
 \put (250,187){\circle*{25}}
 \put (  0,274){\circle*{25}}
 \put (100,274){\circle*{25}}
 \put (200,274){\circle*{25}}
 \put (300,274){\circle*{25}}
 \put (150,361){\circle*{25}}

 \color{black}
 \put (150,13){\circle*{25}}
 \put (150,361){\circle*{25}}
 \put (  0,274){\circle*{25}}
 \put (300,100){\circle*{25}}
\end{picture}}\right)
+V_{2}\left(\ket{%
%
\setlength{\unitlength}{3947sp}
\begin{picture}(320,230)(-10,160)


 \color{magenta}
 \put (10,100){\circle*{1}}
 \put (20,100){\circle*{1}}
 \put (30,100){\circle*{1}}
 \put (40,100){\circle*{1}}
 \put (50,100){\circle*{1}}
 \put (60,100){\circle*{1}}
 \put (70,100){\circle*{1}}
 \put (80,100){\circle*{1}}
 \put (90,100){\circle*{1}}
 \put (95,100){\circle*{1}}

 \put (210,274){\circle*{1}}
 \put (220,274){\circle*{1}}
 \put (230,274){\circle*{1}}
 \put (240,274){\circle*{1}}
 \put (250,274){\circle*{1}}
 \put (260,274){\circle*{1}}
 \put (270,274){\circle*{1}}
 \put (280,274){\circle*{1}}
 \put (290,274){\circle*{1}}
 \put (295,274){\circle*{1}}

 \put ( 50,187){\circle*{1}}
 \put ( 55,197){\circle*{1}}
 \put ( 60,207){\circle*{1}}
 \put ( 65,217){\circle*{1}}
 \put ( 70,227){\circle*{1}}
 \put ( 75,237){\circle*{1}}
 \put ( 80,247){\circle*{1}}
 \put ( 85,257){\circle*{1}}
 \put ( 90,267){\circle*{1}}
 \put ( 95,274){\circle*{1}}

 \put (200,100){\circle*{1}}
 \put (205,110){\circle*{1}}
 \put (210,120){\circle*{1}}
 \put (215,130){\circle*{1}}
 \put (220,140){\circle*{1}}
 \put (225,150){\circle*{1}}
 \put (230,160){\circle*{1}}
 \put (235,170){\circle*{1}}
 \put (240,180){\circle*{1}}
 \put (245,190){\circle*{1}}

 \color{magenta}
 \put (150,13){\circle*{25}}
 \put (  0,100){\circle*{25}}
 \put (100,100){\circle*{25}}
 \put (200,100){\circle*{25}}
 \put (300,100){\circle*{25}}
 \put ( 50,187){\circle*{25}}
 \put (250,187){\circle*{25}}
 \put (  0,274){\circle*{25}}
 \put (100,274){\circle*{25}}
 \put (200,274){\circle*{25}}
 \put (300,274){\circle*{25}}
 \put (150,361){\circle*{25}}

 \color{black}
 \put (150,13){\circle*{25}}
 \put (150,361){\circle*{25}}
 \put (  0,274){\circle*{25}}
 \put (300,100){\circle*{25}}
\end{picture}}\bra{%
%
\setlength{\unitlength}{3947sp}
\begin{picture}(320,230)(-10,160)


 \color{magenta}
 \put (10,100){\circle*{1}}
 \put (20,100){\circle*{1}}
 \put (30,100){\circle*{1}}
 \put (40,100){\circle*{1}}
 \put (50,100){\circle*{1}}
 \put (60,100){\circle*{1}}
 \put (70,100){\circle*{1}}
 \put (80,100){\circle*{1}}
 \put (90,100){\circle*{1}}
 \put (95,100){\circle*{1}}

 \put (210,274){\circle*{1}}
 \put (220,274){\circle*{1}}
 \put (230,274){\circle*{1}}
 \put (240,274){\circle*{1}}
 \put (250,274){\circle*{1}}
 \put (260,274){\circle*{1}}
 \put (270,274){\circle*{1}}
 \put (280,274){\circle*{1}}
 \put (290,274){\circle*{1}}
 \put (295,274){\circle*{1}}

 \put ( 50,187){\circle*{1}}
 \put ( 55,197){\circle*{1}}
 \put ( 60,207){\circle*{1}}
 \put ( 65,217){\circle*{1}}
 \put ( 70,227){\circle*{1}}
 \put ( 75,237){\circle*{1}}
 \put ( 80,247){\circle*{1}}
 \put ( 85,257){\circle*{1}}
 \put ( 90,267){\circle*{1}}
 \put ( 95,274){\circle*{1}}

 \put (200,100){\circle*{1}}
 \put (205,110){\circle*{1}}
 \put (210,120){\circle*{1}}
 \put (215,130){\circle*{1}}
 \put (220,140){\circle*{1}}
 \put (225,150){\circle*{1}}
 \put (230,160){\circle*{1}}
 \put (235,170){\circle*{1}}
 \put (240,180){\circle*{1}}
 \put (245,190){\circle*{1}}

 \color{magenta}
 \put (150,13){\circle*{25}}
 \put (  0,100){\circle*{25}}
 \put (100,100){\circle*{25}}
 \put (200,100){\circle*{25}}
 \put (300,100){\circle*{25}}
 \put ( 50,187){\circle*{25}}
 \put (250,187){\circle*{25}}
 \put (  0,274){\circle*{25}}
 \put (100,274){\circle*{25}}
 \put (200,274){\circle*{25}}
 \put (300,274){\circle*{25}}
 \put (150,361){\circle*{25}}

 \color{black}
 \put (150,13){\circle*{25}}
 \put (150,361){\circle*{25}}
 \put (  0,274){\circle*{25}}
 \put (300,100){\circle*{25}}
\end{picture}} + \ket{%
%
\setlength{\unitlength}{3947sp}
\begin{picture}(320,230)(-10,160)


 \color{magenta}
 \put (110,100){\circle*{1}}
 \put (120,100){\circle*{1}}
 \put (130,100){\circle*{1}}
 \put (140,100){\circle*{1}}
 \put (150,100){\circle*{1}}
 \put (160,100){\circle*{1}}
 \put (170,100){\circle*{1}}
 \put (180,100){\circle*{1}}
 \put (190,100){\circle*{1}}
 \put (195,100){\circle*{1}}

 \put (110,274){\circle*{1}}
 \put (120,274){\circle*{1}}
 \put (130,274){\circle*{1}}
 \put (140,274){\circle*{1}}
 \put (150,274){\circle*{1}}
 \put (160,274){\circle*{1}}
 \put (170,274){\circle*{1}}
 \put (180,274){\circle*{1}}
 \put (190,274){\circle*{1}}
 \put (195,274){\circle*{1}}

 \put (  0,100){\circle*{1}}
 \put (  5,110){\circle*{1}}
 \put ( 10,120){\circle*{1}}
 \put ( 15,130){\circle*{1}}
 \put ( 20,140){\circle*{1}}
 \put ( 25,150){\circle*{1}}
 \put ( 30,160){\circle*{1}}
 \put ( 35,170){\circle*{1}}
 \put ( 40,180){\circle*{1}}
 \put ( 45,187){\circle*{1}}

 \put (250,187){\circle*{1}}
 \put (255,197){\circle*{1}}
 \put (260,207){\circle*{1}}
 \put (265,217){\circle*{1}}
 \put (270,227){\circle*{1}}
 \put (275,237){\circle*{1}}
 \put (280,247){\circle*{1}}
 \put (285,257){\circle*{1}}
 \put (290,267){\circle*{1}}
 \put (295,277){\circle*{1}}

 \color{magenta}
 \put (150,13){\circle*{25}}
 \put (  0,100){\circle*{25}}
 \put (100,100){\circle*{25}}
 \put (200,100){\circle*{25}}
 \put (300,100){\circle*{25}}
 \put ( 50,187){\circle*{25}}
 \put (250,187){\circle*{25}}
 \put (  0,274){\circle*{25}}
 \put (100,274){\circle*{25}}
 \put (200,274){\circle*{25}}
 \put (300,274){\circle*{25}}
 \put (150,361){\circle*{25}}

 \color{black}
 \put (150,13){\circle*{25}}
 \put (150,361){\circle*{25}}
 \put (  0,274){\circle*{25}}
 \put (300,100){\circle*{25}}
\end{picture}}\bra{%
%
\setlength{\unitlength}{3947sp}
\begin{picture}(320,230)(-10,160)


 \color{magenta}
 \put (110,100){\circle*{1}}
 \put (120,100){\circle*{1}}
 \put (130,100){\circle*{1}}
 \put (140,100){\circle*{1}}
 \put (150,100){\circle*{1}}
 \put (160,100){\circle*{1}}
 \put (170,100){\circle*{1}}
 \put (180,100){\circle*{1}}
 \put (190,100){\circle*{1}}
 \put (195,100){\circle*{1}}

 \put (110,274){\circle*{1}}
 \put (120,274){\circle*{1}}
 \put (130,274){\circle*{1}}
 \put (140,274){\circle*{1}}
 \put (150,274){\circle*{1}}
 \put (160,274){\circle*{1}}
 \put (170,274){\circle*{1}}
 \put (180,274){\circle*{1}}
 \put (190,274){\circle*{1}}
 \put (195,274){\circle*{1}}

 \put (  0,100){\circle*{1}}
 \put (  5,110){\circle*{1}}
 \put ( 10,120){\circle*{1}}
 \put ( 15,130){\circle*{1}}
 \put ( 20,140){\circle*{1}}
 \put ( 25,150){\circle*{1}}
 \put ( 30,160){\circle*{1}}
 \put ( 35,170){\circle*{1}}
 \put ( 40,180){\circle*{1}}
 \put ( 45,187){\circle*{1}}

 \put (250,187){\circle*{1}}
 \put (255,197){\circle*{1}}
 \put (260,207){\circle*{1}}
 \put (265,217){\circle*{1}}
 \put (270,227){\circle*{1}}
 \put (275,237){\circle*{1}}
 \put (280,247){\circle*{1}}
 \put (285,257){\circle*{1}}
 \put (290,267){\circle*{1}}
 \put (295,277){\circle*{1}}

 \color{magenta}
 \put (150,13){\circle*{25}}
 \put (  0,100){\circle*{25}}
 \put (100,100){\circle*{25}}
 \put (200,100){\circle*{25}}
 \put (300,100){\circle*{25}}
 \put ( 50,187){\circle*{25}}
 \put (250,187){\circle*{25}}
 \put (  0,274){\circle*{25}}
 \put (100,274){\circle*{25}}
 \put (200,274){\circle*{25}}
 \put (300,274){\circle*{25}}
 \put (150,361){\circle*{25}}

 \color{black}
 \put (150,13){\circle*{25}}
 \put (150,361){\circle*{25}}
 \put (  0,274){\circle*{25}}
 \put (300,100){\circle*{25}}
\end{picture}}\right)\Bigg]\nonumber
\\ 
&&+ \;\ldots\; -t_{32}\left(\ket{%
%
\setlength{\unitlength}{3947sp}
\begin{picture}(320,230)(-10,160)


 \color{magenta}
 \put (10,100){\circle*{1}}
 \put (20,100){\circle*{1}}
 \put (30,100){\circle*{1}}
 \put (40,100){\circle*{1}}
 \put (50,100){\circle*{1}}
 \put (60,100){\circle*{1}}
 \put (70,100){\circle*{1}}
 \put (80,100){\circle*{1}}
 \put (90,100){\circle*{1}}
 \put (95,100){\circle*{1}}

 \put (210,274){\circle*{1}}
 \put (220,274){\circle*{1}}
 \put (230,274){\circle*{1}}
 \put (240,274){\circle*{1}}
 \put (250,274){\circle*{1}}
 \put (260,274){\circle*{1}}
 \put (270,274){\circle*{1}}
 \put (280,274){\circle*{1}}
 \put (290,274){\circle*{1}}
 \put (295,274){\circle*{1}}

 \put (100,274){\circle*{1}}
 \put (105,284){\circle*{1}}
 \put (110,294){\circle*{1}}
 \put (115,304){\circle*{1}}
 \put (120,314){\circle*{1}}
 \put (125,324){\circle*{1}}
 \put (130,334){\circle*{1}}
 \put (135,344){\circle*{1}}
 \put (140,354){\circle*{1}}
 \put (145,364){\circle*{1}}

 \put (50,187){\circle*{1}}
 \put (45,197){\circle*{1}}
 \put (40,207){\circle*{1}}
 \put (35,217){\circle*{1}}
 \put (30,227){\circle*{1}}
 \put (25,237){\circle*{1}}
 \put (20,247){\circle*{1}}
 \put (15,257){\circle*{1}}
 \put (10,267){\circle*{1}}
 \put ( 5,277){\circle*{1}}

 \put (150, 13){\circle*{1}}
 \put (155, 23){\circle*{1}}
 \put (160, 33){\circle*{1}}
 \put (165, 43){\circle*{1}}
 \put (170, 53){\circle*{1}}
 \put (175, 63){\circle*{1}}
 \put (180, 73){\circle*{1}}
 \put (185, 83){\circle*{1}}
 \put (190, 93){\circle*{1}}
 \put (195, 103){\circle*{1}}

 \put (300,100){\circle*{1}}
 \put (295,110){\circle*{1}}
 \put (290,120){\circle*{1}}
 \put (285,130){\circle*{1}}
 \put (280,140){\circle*{1}}
 \put (275,150){\circle*{1}}
 \put (270,160){\circle*{1}}
 \put (265,170){\circle*{1}}
 \put (260,180){\circle*{1}}
 \put (255,190){\circle*{1}}

 \color{magenta}
 \put (150,13){\circle*{25}}
 \put (  0,100){\circle*{25}}
 \put (100,100){\circle*{25}}
 \put (200,100){\circle*{25}}
 \put (300,100){\circle*{25}}
 \put ( 50,187){\circle*{25}}
 \put (250,187){\circle*{25}}
 \put (  0,274){\circle*{25}}
 \put (100,274){\circle*{25}}
 \put (200,274){\circle*{25}}
 \put (300,274){\circle*{25}}
 \put (150,361){\circle*{25}}

\end{picture}}\bra{%
%
\setlength{\unitlength}{3947sp}
\begin{picture}(320,230)(-10,160)


 \color{magenta}
 \put (210,100){\circle*{1}}
 \put (220,100){\circle*{1}}
 \put (230,100){\circle*{1}}
 \put (240,100){\circle*{1}}
 \put (250,100){\circle*{1}}
 \put (260,100){\circle*{1}}
 \put (270,100){\circle*{1}}
 \put (280,100){\circle*{1}}
 \put (290,100){\circle*{1}}
 \put (295,100){\circle*{1}}

 \put (10,274){\circle*{1}}
 \put (20,274){\circle*{1}}
 \put (30,274){\circle*{1}}
 \put (40,274){\circle*{1}}
 \put (50,274){\circle*{1}}
 \put (60,274){\circle*{1}}
 \put (70,274){\circle*{1}}
 \put (80,274){\circle*{1}}
 \put (90,274){\circle*{1}}
 \put (95,274){\circle*{1}}

 \put (250,187){\circle*{1}}
 \put (255,197){\circle*{1}}
 \put (260,207){\circle*{1}}
 \put (265,217){\circle*{1}}
 \put (270,227){\circle*{1}}
 \put (275,237){\circle*{1}}
 \put (280,247){\circle*{1}}
 \put (285,257){\circle*{1}}
 \put (290,267){\circle*{1}}
 \put (295,277){\circle*{1}}

 \put ( 0,100){\circle*{1}}
 \put ( 5,110){\circle*{1}}
 \put (10,120){\circle*{1}}
 \put (15,130){\circle*{1}}
 \put (20,140){\circle*{1}}
 \put (25,150){\circle*{1}}
 \put (30,160){\circle*{1}}
 \put (35,170){\circle*{1}}
 \put (45,180){\circle*{1}}
 \put (50,190){\circle*{1}}

 \put (150, 13){\circle*{1}}
 \put (145, 23){\circle*{1}}
 \put (140, 33){\circle*{1}}
 \put (135, 43){\circle*{1}}
 \put (130, 53){\circle*{1}}
 \put (125, 63){\circle*{1}}
 \put (120, 73){\circle*{1}}
 \put (115, 83){\circle*{1}}
 \put (110, 93){\circle*{1}}
 \put (105, 103){\circle*{1}}

 \put (200,274){\circle*{1}}
 \put (195,284){\circle*{1}}
 \put (190,294){\circle*{1}}
 \put (185,304){\circle*{1}}
 \put (180,314){\circle*{1}}
 \put (175,324){\circle*{1}}
 \put (170,334){\circle*{1}}
 \put (165,344){\circle*{1}}
 \put (160,354){\circle*{1}}
 \put (155,364){\circle*{1}}

 \color{magenta}
 \put (150,13){\circle*{25}}
 \put (  0,100){\circle*{25}}
 \put (100,100){\circle*{25}}
 \put (200,100){\circle*{25}}
 \put (300,100){\circle*{25}}
 \put ( 50,187){\circle*{25}}
 \put (250,187){\circle*{25}}
 \put (  0,274){\circle*{25}}
 \put (100,274){\circle*{25}}
 \put (200,274){\circle*{25}}
 \put (300,274){\circle*{25}}
 \put (150,361){\circle*{25}}

\end{picture}} + \ket{%
%
\setlength{\unitlength}{3947sp}
\begin{picture}(320,230)(-10,160)


 \color{magenta}
 \put (210,100){\circle*{1}}
 \put (220,100){\circle*{1}}
 \put (230,100){\circle*{1}}
 \put (240,100){\circle*{1}}
 \put (250,100){\circle*{1}}
 \put (260,100){\circle*{1}}
 \put (270,100){\circle*{1}}
 \put (280,100){\circle*{1}}
 \put (290,100){\circle*{1}}
 \put (295,100){\circle*{1}}

 \put (10,274){\circle*{1}}
 \put (20,274){\circle*{1}}
 \put (30,274){\circle*{1}}
 \put (40,274){\circle*{1}}
 \put (50,274){\circle*{1}}
 \put (60,274){\circle*{1}}
 \put (70,274){\circle*{1}}
 \put (80,274){\circle*{1}}
 \put (90,274){\circle*{1}}
 \put (95,274){\circle*{1}}

 \put (250,187){\circle*{1}}
 \put (255,197){\circle*{1}}
 \put (260,207){\circle*{1}}
 \put (265,217){\circle*{1}}
 \put (270,227){\circle*{1}}
 \put (275,237){\circle*{1}}
 \put (280,247){\circle*{1}}
 \put (285,257){\circle*{1}}
 \put (290,267){\circle*{1}}
 \put (295,277){\circle*{1}}

 \put ( 0,100){\circle*{1}}
 \put ( 5,110){\circle*{1}}
 \put (10,120){\circle*{1}}
 \put (15,130){\circle*{1}}
 \put (20,140){\circle*{1}}
 \put (25,150){\circle*{1}}
 \put (30,160){\circle*{1}}
 \put (35,170){\circle*{1}}
 \put (45,180){\circle*{1}}
 \put (50,190){\circle*{1}}

 \put (150, 13){\circle*{1}}
 \put (145, 23){\circle*{1}}
 \put (140, 33){\circle*{1}}
 \put (135, 43){\circle*{1}}
 \put (130, 53){\circle*{1}}
 \put (125, 63){\circle*{1}}
 \put (120, 73){\circle*{1}}
 \put (115, 83){\circle*{1}}
 \put (110, 93){\circle*{1}}
 \put (105, 103){\circle*{1}}

 \put (200,274){\circle*{1}}
 \put (195,284){\circle*{1}}
 \put (190,294){\circle*{1}}
 \put (185,304){\circle*{1}}
 \put (180,314){\circle*{1}}
 \put (175,324){\circle*{1}}
 \put (170,334){\circle*{1}}
 \put (165,344){\circle*{1}}
 \put (160,354){\circle*{1}}
 \put (155,364){\circle*{1}}

 \color{magenta}
 \put (150,13){\circle*{25}}
 \put (  0,100){\circle*{25}}
 \put (100,100){\circle*{25}}
 \put (200,100){\circle*{25}}
 \put (300,100){\circle*{25}}
 \put ( 50,187){\circle*{25}}
 \put (250,187){\circle*{25}}
 \put (  0,274){\circle*{25}}
 \put (100,274){\circle*{25}}
 \put (200,274){\circle*{25}}
 \put (300,274){\circle*{25}}
 \put (150,361){\circle*{25}}

\end{picture}}\bra{%
%
\setlength{\unitlength}{3947sp}
\begin{picture}(320,230)(-10,160)


 \color{magenta}
 \put (10,100){\circle*{1}}
 \put (20,100){\circle*{1}}
 \put (30,100){\circle*{1}}
 \put (40,100){\circle*{1}}
 \put (50,100){\circle*{1}}
 \put (60,100){\circle*{1}}
 \put (70,100){\circle*{1}}
 \put (80,100){\circle*{1}}
 \put (90,100){\circle*{1}}
 \put (95,100){\circle*{1}}

 \put (210,274){\circle*{1}}
 \put (220,274){\circle*{1}}
 \put (230,274){\circle*{1}}
 \put (240,274){\circle*{1}}
 \put (250,274){\circle*{1}}
 \put (260,274){\circle*{1}}
 \put (270,274){\circle*{1}}
 \put (280,274){\circle*{1}}
 \put (290,274){\circle*{1}}
 \put (295,274){\circle*{1}}

 \put (100,274){\circle*{1}}
 \put (105,284){\circle*{1}}
 \put (110,294){\circle*{1}}
 \put (115,304){\circle*{1}}
 \put (120,314){\circle*{1}}
 \put (125,324){\circle*{1}}
 \put (130,334){\circle*{1}}
 \put (135,344){\circle*{1}}
 \put (140,354){\circle*{1}}
 \put (145,364){\circle*{1}}

 \put (50,187){\circle*{1}}
 \put (45,197){\circle*{1}}
 \put (40,207){\circle*{1}}
 \put (35,217){\circle*{1}}
 \put (30,227){\circle*{1}}
 \put (25,237){\circle*{1}}
 \put (20,247){\circle*{1}}
 \put (15,257){\circle*{1}}
 \put (10,267){\circle*{1}}
 \put ( 5,277){\circle*{1}}

 \put (150, 13){\circle*{1}}
 \put (155, 23){\circle*{1}}
 \put (160, 33){\circle*{1}}
 \put (165, 43){\circle*{1}}
 \put (170, 53){\circle*{1}}
 \put (175, 63){\circle*{1}}
 \put (180, 73){\circle*{1}}
 \put (185, 83){\circle*{1}}
 \put (190, 93){\circle*{1}}
 \put (195, 103){\circle*{1}}

 \put (300,100){\circle*{1}}
 \put (295,110){\circle*{1}}
 \put (290,120){\circle*{1}}
 \put (285,130){\circle*{1}}
 \put (280,140){\circle*{1}}
 \put (275,150){\circle*{1}}
 \put (270,160){\circle*{1}}
 \put (265,170){\circle*{1}}
 \put (260,180){\circle*{1}}
 \put (255,190){\circle*{1}}

 \color{magenta}
 \put (150,13){\circle*{25}}
 \put (  0,100){\circle*{25}}
 \put (100,100){\circle*{25}}
 \put (200,100){\circle*{25}}
 \put (300,100){\circle*{25}}
 \put ( 50,187){\circle*{25}}
 \put (250,187){\circle*{25}}
 \put (  0,274){\circle*{25}}
 \put (100,274){\circle*{25}}
 \put (200,274){\circle*{25}}
 \put (300,274){\circle*{25}}
 \put (150,361){\circle*{25}}

\end{picture}}\right)
+V_{32}\left(\ket{%
%
\setlength{\unitlength}{3947sp}
\begin{picture}(320,230)(-10,160)


 \color{magenta}
 \put (10,100){\circle*{1}}
 \put (20,100){\circle*{1}}
 \put (30,100){\circle*{1}}
 \put (40,100){\circle*{1}}
 \put (50,100){\circle*{1}}
 \put (60,100){\circle*{1}}
 \put (70,100){\circle*{1}}
 \put (80,100){\circle*{1}}
 \put (90,100){\circle*{1}}
 \put (95,100){\circle*{1}}

 \put (210,274){\circle*{1}}
 \put (220,274){\circle*{1}}
 \put (230,274){\circle*{1}}
 \put (240,274){\circle*{1}}
 \put (250,274){\circle*{1}}
 \put (260,274){\circle*{1}}
 \put (270,274){\circle*{1}}
 \put (280,274){\circle*{1}}
 \put (290,274){\circle*{1}}
 \put (295,274){\circle*{1}}

 \put (100,274){\circle*{1}}
 \put (105,284){\circle*{1}}
 \put (110,294){\circle*{1}}
 \put (115,304){\circle*{1}}
 \put (120,314){\circle*{1}}
 \put (125,324){\circle*{1}}
 \put (130,334){\circle*{1}}
 \put (135,344){\circle*{1}}
 \put (140,354){\circle*{1}}
 \put (145,364){\circle*{1}}

 \put (50,187){\circle*{1}}
 \put (45,197){\circle*{1}}
 \put (40,207){\circle*{1}}
 \put (35,217){\circle*{1}}
 \put (30,227){\circle*{1}}
 \put (25,237){\circle*{1}}
 \put (20,247){\circle*{1}}
 \put (15,257){\circle*{1}}
 \put (10,267){\circle*{1}}
 \put ( 5,277){\circle*{1}}

 \put (150, 13){\circle*{1}}
 \put (155, 23){\circle*{1}}
 \put (160, 33){\circle*{1}}
 \put (165, 43){\circle*{1}}
 \put (170, 53){\circle*{1}}
 \put (175, 63){\circle*{1}}
 \put (180, 73){\circle*{1}}
 \put (185, 83){\circle*{1}}
 \put (190, 93){\circle*{1}}
 \put (195, 103){\circle*{1}}

 \put (300,100){\circle*{1}}
 \put (295,110){\circle*{1}}
 \put (290,120){\circle*{1}}
 \put (285,130){\circle*{1}}
 \put (280,140){\circle*{1}}
 \put (275,150){\circle*{1}}
 \put (270,160){\circle*{1}}
 \put (265,170){\circle*{1}}
 \put (260,180){\circle*{1}}
 \put (255,190){\circle*{1}}

 \color{magenta}
 \put (150,13){\circle*{25}}
 \put (  0,100){\circle*{25}}
 \put (100,100){\circle*{25}}
 \put (200,100){\circle*{25}}
 \put (300,100){\circle*{25}}
 \put ( 50,187){\circle*{25}}
 \put (250,187){\circle*{25}}
 \put (  0,274){\circle*{25}}
 \put (100,274){\circle*{25}}
 \put (200,274){\circle*{25}}
 \put (300,274){\circle*{25}}
 \put (150,361){\circle*{25}}

\end{picture}}\bra{%
%
\setlength{\unitlength}{3947sp}
\begin{picture}(320,230)(-10,160)


 \color{magenta}
 \put (10,100){\circle*{1}}
 \put (20,100){\circle*{1}}
 \put (30,100){\circle*{1}}
 \put (40,100){\circle*{1}}
 \put (50,100){\circle*{1}}
 \put (60,100){\circle*{1}}
 \put (70,100){\circle*{1}}
 \put (80,100){\circle*{1}}
 \put (90,100){\circle*{1}}
 \put (95,100){\circle*{1}}

 \put (210,274){\circle*{1}}
 \put (220,274){\circle*{1}}
 \put (230,274){\circle*{1}}
 \put (240,274){\circle*{1}}
 \put (250,274){\circle*{1}}
 \put (260,274){\circle*{1}}
 \put (270,274){\circle*{1}}
 \put (280,274){\circle*{1}}
 \put (290,274){\circle*{1}}
 \put (295,274){\circle*{1}}

 \put (100,274){\circle*{1}}
 \put (105,284){\circle*{1}}
 \put (110,294){\circle*{1}}
 \put (115,304){\circle*{1}}
 \put (120,314){\circle*{1}}
 \put (125,324){\circle*{1}}
 \put (130,334){\circle*{1}}
 \put (135,344){\circle*{1}}
 \put (140,354){\circle*{1}}
 \put (145,364){\circle*{1}}

 \put (50,187){\circle*{1}}
 \put (45,197){\circle*{1}}
 \put (40,207){\circle*{1}}
 \put (35,217){\circle*{1}}
 \put (30,227){\circle*{1}}
 \put (25,237){\circle*{1}}
 \put (20,247){\circle*{1}}
 \put (15,257){\circle*{1}}
 \put (10,267){\circle*{1}}
 \put ( 5,277){\circle*{1}}

 \put (150, 13){\circle*{1}}
 \put (155, 23){\circle*{1}}
 \put (160, 33){\circle*{1}}
 \put (165, 43){\circle*{1}}
 \put (170, 53){\circle*{1}}
 \put (175, 63){\circle*{1}}
 \put (180, 73){\circle*{1}}
 \put (185, 83){\circle*{1}}
 \put (190, 93){\circle*{1}}
 \put (195, 103){\circle*{1}}

 \put (300,100){\circle*{1}}
 \put (295,110){\circle*{1}}
 \put (290,120){\circle*{1}}
 \put (285,130){\circle*{1}}
 \put (280,140){\circle*{1}}
 \put (275,150){\circle*{1}}
 \put (270,160){\circle*{1}}
 \put (265,170){\circle*{1}}
 \put (260,180){\circle*{1}}
 \put (255,190){\circle*{1}}

 \color{magenta}
 \put (150,13){\circle*{25}}
 \put (  0,100){\circle*{25}}
 \put (100,100){\circle*{25}}
 \put (200,100){\circle*{25}}
 \put (300,100){\circle*{25}}
 \put ( 50,187){\circle*{25}}
 \put (250,187){\circle*{25}}
 \put (  0,274){\circle*{25}}
 \put (100,274){\circle*{25}}
 \put (200,274){\circle*{25}}
 \put (300,274){\circle*{25}}
 \put (150,361){\circle*{25}}

\end{picture}} + \ket{%
%
\setlength{\unitlength}{3947sp}
\begin{picture}(320,230)(-10,160)


 \color{magenta}
 \put (210,100){\circle*{1}}
 \put (220,100){\circle*{1}}
 \put (230,100){\circle*{1}}
 \put (240,100){\circle*{1}}
 \put (250,100){\circle*{1}}
 \put (260,100){\circle*{1}}
 \put (270,100){\circle*{1}}
 \put (280,100){\circle*{1}}
 \put (290,100){\circle*{1}}
 \put (295,100){\circle*{1}}

 \put (10,274){\circle*{1}}
 \put (20,274){\circle*{1}}
 \put (30,274){\circle*{1}}
 \put (40,274){\circle*{1}}
 \put (50,274){\circle*{1}}
 \put (60,274){\circle*{1}}
 \put (70,274){\circle*{1}}
 \put (80,274){\circle*{1}}
 \put (90,274){\circle*{1}}
 \put (95,274){\circle*{1}}

 \put (250,187){\circle*{1}}
 \put (255,197){\circle*{1}}
 \put (260,207){\circle*{1}}
 \put (265,217){\circle*{1}}
 \put (270,227){\circle*{1}}
 \put (275,237){\circle*{1}}
 \put (280,247){\circle*{1}}
 \put (285,257){\circle*{1}}
 \put (290,267){\circle*{1}}
 \put (295,277){\circle*{1}}

 \put ( 0,100){\circle*{1}}
 \put ( 5,110){\circle*{1}}
 \put (10,120){\circle*{1}}
 \put (15,130){\circle*{1}}
 \put (20,140){\circle*{1}}
 \put (25,150){\circle*{1}}
 \put (30,160){\circle*{1}}
 \put (35,170){\circle*{1}}
 \put (45,180){\circle*{1}}
 \put (50,190){\circle*{1}}

 \put (150, 13){\circle*{1}}
 \put (145, 23){\circle*{1}}
 \put (140, 33){\circle*{1}}
 \put (135, 43){\circle*{1}}
 \put (130, 53){\circle*{1}}
 \put (125, 63){\circle*{1}}
 \put (120, 73){\circle*{1}}
 \put (115, 83){\circle*{1}}
 \put (110, 93){\circle*{1}}
 \put (105, 103){\circle*{1}}

 \put (200,274){\circle*{1}}
 \put (195,284){\circle*{1}}
 \put (190,294){\circle*{1}}
 \put (185,304){\circle*{1}}
 \put (180,314){\circle*{1}}
 \put (175,324){\circle*{1}}
 \put (170,334){\circle*{1}}
 \put (165,344){\circle*{1}}
 \put (160,354){\circle*{1}}
 \put (155,364){\circle*{1}}

 \color{magenta}
 \put (150,13){\circle*{25}}
 \put (  0,100){\circle*{25}}
 \put (100,100){\circle*{25}}
 \put (200,100){\circle*{25}}
 \put (300,100){\circle*{25}}
 \put ( 50,187){\circle*{25}}
 \put (250,187){\circle*{25}}
 \put (  0,274){\circle*{25}}
 \put (100,274){\circle*{25}}
 \put (200,274){\circle*{25}}
 \put (300,274){\circle*{25}}
 \put (150,361){\circle*{25}}

\end{picture}}\bra{%
%
\setlength{\unitlength}{3947sp}
\begin{picture}(320,230)(-10,160)


 \color{magenta}
 \put (210,100){\circle*{1}}
 \put (220,100){\circle*{1}}
 \put (230,100){\circle*{1}}
 \put (240,100){\circle*{1}}
 \put (250,100){\circle*{1}}
 \put (260,100){\circle*{1}}
 \put (270,100){\circle*{1}}
 \put (280,100){\circle*{1}}
 \put (290,100){\circle*{1}}
 \put (295,100){\circle*{1}}

 \put (10,274){\circle*{1}}
 \put (20,274){\circle*{1}}
 \put (30,274){\circle*{1}}
 \put (40,274){\circle*{1}}
 \put (50,274){\circle*{1}}
 \put (60,274){\circle*{1}}
 \put (70,274){\circle*{1}}
 \put (80,274){\circle*{1}}
 \put (90,274){\circle*{1}}
 \put (95,274){\circle*{1}}

 \put (250,187){\circle*{1}}
 \put (255,197){\circle*{1}}
 \put (260,207){\circle*{1}}
 \put (265,217){\circle*{1}}
 \put (270,227){\circle*{1}}
 \put (275,237){\circle*{1}}
 \put (280,247){\circle*{1}}
 \put (285,257){\circle*{1}}
 \put (290,267){\circle*{1}}
 \put (295,277){\circle*{1}}

 \put ( 0,100){\circle*{1}}
 \put ( 5,110){\circle*{1}}
 \put (10,120){\circle*{1}}
 \put (15,130){\circle*{1}}
 \put (20,140){\circle*{1}}
 \put (25,150){\circle*{1}}
 \put (30,160){\circle*{1}}
 \put (35,170){\circle*{1}}
 \put (45,180){\circle*{1}}
 \put (50,190){\circle*{1}}

 \put (150, 13){\circle*{1}}
 \put (145, 23){\circle*{1}}
 \put (140, 33){\circle*{1}}
 \put (135, 43){\circle*{1}}
 \put (130, 53){\circle*{1}}
 \put (125, 63){\circle*{1}}
 \put (120, 73){\circle*{1}}
 \put (115, 83){\circle*{1}}
 \put (110, 93){\circle*{1}}
 \put (105, 103){\circle*{1}}

 \put (200,274){\circle*{1}}
 \put (195,284){\circle*{1}}
 \put (190,294){\circle*{1}}
 \put (185,304){\circle*{1}}
 \put (180,314){\circle*{1}}
 \put (175,324){\circle*{1}}
 \put (170,334){\circle*{1}}
 \put (165,344){\circle*{1}}
 \put (160,354){\circle*{1}}
 \put (155,364){\circle*{1}}

 \color{magenta}
 \put (150,13){\circle*{25}}
 \put (  0,100){\circle*{25}}
 \put (100,100){\circle*{25}}
 \put (200,100){\circle*{25}}
 \put (300,100){\circle*{25}}
 \put ( 50,187){\circle*{25}}
 \put (250,187){\circle*{25}}
 \put (  0,274){\circle*{25}}
 \put (100,274){\circle*{25}}
 \put (200,274){\circle*{25}}
 \put (300,274){\circle*{25}}
 \put (150,361){\circle*{25}}

\end{picture}}
\right)\Bigg]\;. 
\end{eqnarray}
\end{widetext}
Magenta bonds indicate occupancy by dimers. 
In the above, we sum over all 12-site star plaquettes of the lattice.  
All kinetic terms execute ``resonance moves'' along one of 32 loops contained within the star, such that 
occupied links alternate along the loop, and the move changes the occupancy along the loop (cf. Table \ref{table3}). 
It is easy to see that each dimer covering results in precisely one such  move being possible per star \cite{misguich}. 
The potential terms associate an energy with the associated loop. 

For toroidal topology, i.e., periodic boundary conditions (PBCs), 
dimer configurations can be classified according to winding numbers $W_{x}$ and $W_{y}$.  
Dimer configurations with different winding numbers are thought of as belonging to different 
topological sectors and cannot be connected by local resonance moves of dimers. 
To determine the winding number $W_{x}$ ($W_{y}$) one considers a horizontal (vertical) line 
around the torus which intersects the links. 
$W_{x}$ ($W_{y}$) is then the parity of the number of dimers intersected. 
\begin{table}[!b]
\begin{tabular}{|c|c|c|c|}
  \hline \hline
$\color{white}{{\Bigg|}}%
%
\setlength{\unitlength}{3947sp}
\begin{picture}(740,660)(-20,320)

 \color{magenta}
 \put (100,374){\circle*{1}}
 \put (105,384){\circle*{1}}
 \put (110,394){\circle*{1}}
 \put (115,404){\circle*{1}}
 \put (120,414){\circle*{1}}
 \put (125,424){\circle*{1}}
 \put (130,434){\circle*{1}}
 \put (135,444){\circle*{1}}
 \put (140,454){\circle*{1}}
 \put (145,464){\circle*{1}}
 \put (150,474){\circle*{1}}
 \put (155,484){\circle*{1}}
 \put (160,494){\circle*{1}}
 \put (165,504){\circle*{1}}
 \put (170,514){\circle*{1}}
 \put (175,524){\circle*{1}}
 \put (180,534){\circle*{1}}
 \put (185,544){\circle*{1}}
 \put (190,554){\circle*{1}}
 \put (195,564){\circle*{1}}

 \put (500,374){\circle*{1}}
 \put (495,384){\circle*{1}}
 \put (490,394){\circle*{1}}
 \put (485,404){\circle*{1}}
 \put (480,414){\circle*{1}}
 \put (475,424){\circle*{1}}
 \put (470,434){\circle*{1}}
 \put (465,444){\circle*{1}}
 \put (460,454){\circle*{1}}
 \put (455,464){\circle*{1}}
 \put (450,474){\circle*{1}}
 \put (445,484){\circle*{1}}
 \put (440,494){\circle*{1}}
 \put (435,504){\circle*{1}}
 \put (430,514){\circle*{1}}
 \put (425,524){\circle*{1}}
 \put (420,534){\circle*{1}}
 \put (415,544){\circle*{1}}
 \put (410,554){\circle*{1}}
 \put (405,564){\circle*{1}}

 \put (220,548){\circle*{10}}
 \put (230,548){\circle*{10}}
 \put (240,548){\circle*{10}}
 \put (250,548){\circle*{10}}
 \put (260,548){\circle*{10}}
 \put (270,548){\circle*{10}}
 \put (280,548){\circle*{10}}
 \put (290,548){\circle*{10}}
 \put (300,548){\circle*{10}}
 \put (310,548){\circle*{10}}
 \put (320,548){\circle*{10}}
 \put (330,548){\circle*{10}}
 \put (340,548){\circle*{10}}
 \put (350,548){\circle*{10}}
 \put (360,548){\circle*{10}}
 \put (370,548){\circle*{10}}
 \put (380,548){\circle*{10}}
 \put (390,548){\circle*{10}}
 \put (400,548){\circle*{10}}

 \put (220,200){\circle*{10}}
 \put (230,200){\circle*{10}}
 \put (240,200){\circle*{10}}
 \put (250,200){\circle*{10}}
 \put (260,200){\circle*{10}}
 \put (270,200){\circle*{10}}
 \put (280,200){\circle*{10}}
 \put (290,200){\circle*{10}}
 \put (300,200){\circle*{10}}
 \put (310,200){\circle*{10}}
 \put (320,200){\circle*{10}}
 \put (330,200){\circle*{10}}
 \put (340,200){\circle*{10}}
 \put (350,200){\circle*{10}}
 \put (360,200){\circle*{10}}
 \put (370,200){\circle*{10}}
 \put (380,200){\circle*{10}}
 \put (390,200){\circle*{10}}
 \put (400,200){\circle*{10}} 

 \color{magenta}
 \put (210,200){\circle*{10}}
 \put (200,210){\circle*{10}}
 \put (190,220){\circle*{10}}
 \put (180,230){\circle*{10}}
 \put (170,240){\circle*{10}}
 \put (165,250){\circle*{10}}
 \put (160,260){\circle*{10}}
 \put (155,270){\circle*{10}}
 \put (150,280){\circle*{10}}
 \put (145,290){\circle*{10}}
 \put (140,300){\circle*{10}}
 \put (135,310){\circle*{10}}
 \put (130,320){\circle*{10}}
 \put (125,330){\circle*{10}}
 \put (120,340){\circle*{10}}
 \put (115,350){\circle*{10}}
 \put (110,360){\circle*{10}}
 \put (105,370){\circle*{10}}
 \put (100,374){\circle*{10}}
 \put ( 95,380){\circle*{10}}
 
 \color{magenta}
 \put (400,200){\circle*{10}}
 \put (405,210){\circle*{10}}
 \put (410,220){\circle*{10}}
 \put (420,230){\circle*{10}}
 \put (430,240){\circle*{10}}
 \put (435,250){\circle*{10}}
 \put (440,260){\circle*{10}}
 \put (445,270){\circle*{10}}
 \put (450,280){\circle*{10}}
 \put (455,290){\circle*{10}}
 \put (460,300){\circle*{10}}
 \put (465,310){\circle*{10}}
 \put (470,320){\circle*{10}}
 \put (475,330){\circle*{10}}
 \put (480,340){\circle*{10}}
 \put (485,350){\circle*{10}}
 \put (490,360){\circle*{10}}
 \put (495,370){\circle*{10}}
 \put (500,374){\circle*{10}}
 \put (505,380){\circle*{10}}

 \color{magenta}
 \put (300,26){\circle*{50}}
 \put (  0,200){\circle*{50}}
 \put (200,200){\circle*{50}}
 \put (400,200){\circle*{50}}
 \put (600,200){\circle*{50}}
 \put (100,374){\circle*{50}}
 \put (500,374){\circle*{50}}
 \put (  0,548){\circle*{50}}
 \put (200,548){\circle*{50}}
 \put (400,548){\circle*{50}}
 \put (600,548){\circle*{50}}
 \put (300,722){\circle*{50}}

 \color{black}
 \put (  0,548){\circle*{50}}
 \put (600,548){\circle*{50}}
 \put (  0,200){\circle*{50}}
 \put (600,200){\circle*{50}}
 \put (300, 26){\circle*{50}}
 \put (300,722){\circle*{50}}
\end{picture}$ & $\color{white}{{\Bigg|}}%
%
\setlength{\unitlength}{3947sp}
\begin{picture}(740,660)(-20,320)

 \color{magenta}
 \put (  0,174){\circle*{1}}
 \put (  5,184){\circle*{1}}
 \put ( 10,194){\circle*{1}}
 \put ( 15,204){\circle*{1}}
 \put ( 20,214){\circle*{1}}
 \put ( 25,224){\circle*{1}}
 \put ( 30,234){\circle*{1}}
 \put ( 35,244){\circle*{1}}
 \put ( 40,254){\circle*{1}}
 \put ( 45,264){\circle*{1}}
 \put ( 50,274){\circle*{1}}
 \put ( 55,284){\circle*{1}}
 \put ( 60,294){\circle*{1}}
 \put ( 65,304){\circle*{1}}
 \put ( 70,314){\circle*{1}}
 \put ( 75,324){\circle*{1}}
 \put ( 80,334){\circle*{1}}
 \put ( 85,344){\circle*{1}}
 \put ( 90,354){\circle*{1}}
 \put ( 95,364){\circle*{1}}

 \put (100,374){\circle*{1}}
 \put (105,384){\circle*{1}}
 \put (110,394){\circle*{1}}
 \put (115,404){\circle*{1}}
 \put (120,414){\circle*{1}}
 \put (125,424){\circle*{1}}
 \put (130,434){\circle*{1}}
 \put (135,444){\circle*{1}}
 \put (140,454){\circle*{1}}
 \put (145,464){\circle*{1}}
 \put (150,474){\circle*{1}}
 \put (155,484){\circle*{1}}
 \put (160,494){\circle*{1}}
 \put (165,504){\circle*{1}}
 \put (170,514){\circle*{1}}
 \put (175,524){\circle*{1}}
 \put (180,534){\circle*{1}}
 \put (185,544){\circle*{1}}
 \put (190,554){\circle*{1}}
 \put (195,564){\circle*{1}}


 \put (500,374){\circle*{1}}
 \put (505,384){\circle*{1}}
 \put (510,394){\circle*{1}}
 \put (515,404){\circle*{1}}
 \put (520,414){\circle*{1}}
 \put (525,424){\circle*{1}}
 \put (530,434){\circle*{1}}
 \put (535,444){\circle*{1}}
 \put (540,454){\circle*{1}}
 \put (545,464){\circle*{1}}
 \put (550,474){\circle*{1}}
 \put (555,484){\circle*{1}}
 \put (560,494){\circle*{1}}
 \put (565,504){\circle*{1}}
 \put (570,514){\circle*{1}}
 \put (575,524){\circle*{1}}
 \put (580,534){\circle*{1}}
 \put (585,544){\circle*{1}}
 \put (590,554){\circle*{1}}
 \put (595,564){\circle*{1}}

 \put (420,548){\circle*{10}}
 \put (430,548){\circle*{10}}
 \put (440,548){\circle*{10}}
 \put (450,548){\circle*{10}}
 \put (460,548){\circle*{10}}
 \put (470,548){\circle*{10}}
 \put (480,548){\circle*{10}}
 \put (490,548){\circle*{10}}
 \put (500,548){\circle*{10}}
 \put (510,548){\circle*{10}}
 \put (520,548){\circle*{10}}
 \put (530,548){\circle*{10}}
 \put (540,548){\circle*{10}}
 \put (550,548){\circle*{10}}
 \put (560,548){\circle*{10}}
 \put (570,548){\circle*{10}}
 \put (580,548){\circle*{10}}
 \put (590,548){\circle*{10}}
 \put (500,548){\circle*{10}} 

 \put (220,548){\circle*{10}}
 \put (230,548){\circle*{10}}
 \put (240,548){\circle*{10}}
 \put (250,548){\circle*{10}}
 \put (260,548){\circle*{10}}
 \put (270,548){\circle*{10}}
 \put (280,548){\circle*{10}}
 \put (290,548){\circle*{10}}
 \put (300,548){\circle*{10}}
 \put (310,548){\circle*{10}}
 \put (320,548){\circle*{10}}
 \put (330,548){\circle*{10}}
 \put (340,548){\circle*{10}}
 \put (350,548){\circle*{10}}
 \put (360,548){\circle*{10}}
 \put (370,548){\circle*{10}}
 \put (380,548){\circle*{10}}
 \put (390,548){\circle*{10}}
 \put (400,548){\circle*{10}}


 \put ( 20,200){\circle*{10}}
 \put ( 30,200){\circle*{10}}
 \put ( 40,200){\circle*{10}}
 \put ( 50,200){\circle*{10}}
 \put ( 60,200){\circle*{10}}
 \put ( 70,200){\circle*{10}}
 \put ( 80,200){\circle*{10}}
 \put ( 90,200){\circle*{10}}
 \put (100,200){\circle*{10}}
 \put (110,200){\circle*{10}}
 \put (120,200){\circle*{10}}
 \put (130,200){\circle*{10}}
 \put (140,200){\circle*{10}}
 \put (150,200){\circle*{10}}
 \put (160,200){\circle*{10}}
 \put (170,200){\circle*{10}}
 \put (180,200){\circle*{10}}
 \put (190,200){\circle*{10}}
 \put (200,200){\circle*{10}} 

 \put (220,200){\circle*{10}}
 \put (230,200){\circle*{10}}
 \put (240,200){\circle*{10}}
 \put (250,200){\circle*{10}}
 \put (260,200){\circle*{10}}
 \put (270,200){\circle*{10}}
 \put (280,200){\circle*{10}}
 \put (290,200){\circle*{10}}
 \put (300,200){\circle*{10}}
 \put (310,200){\circle*{10}}
 \put (320,200){\circle*{10}}
 \put (330,200){\circle*{10}}
 \put (340,200){\circle*{10}}
 \put (350,200){\circle*{10}}
 \put (360,200){\circle*{10}}
 \put (370,200){\circle*{10}}
 \put (380,200){\circle*{10}}
 \put (390,200){\circle*{10}}
 \put (400,200){\circle*{10}} 

 
 \color{magenta}


 \color{magenta}
 \put (400,200){\circle*{10}}
 \put (405,210){\circle*{10}}
 \put (410,220){\circle*{10}}
 \put (420,230){\circle*{10}}
 \put (430,240){\circle*{10}}
 \put (435,250){\circle*{10}}
 \put (440,260){\circle*{10}}
 \put (445,270){\circle*{10}}
 \put (450,280){\circle*{10}}
 \put (455,290){\circle*{10}}
 \put (460,300){\circle*{10}}
 \put (465,310){\circle*{10}}
 \put (470,320){\circle*{10}}
 \put (475,330){\circle*{10}}
 \put (480,340){\circle*{10}}
 \put (485,350){\circle*{10}}
 \put (490,360){\circle*{10}}
 \put (495,370){\circle*{10}}
 \put (500,374){\circle*{10}}
 \put (505,380){\circle*{10}}

 \color{magenta}
 \put (300,26){\circle*{50}}
 \put (  0,200){\circle*{50}}
 \put (200,200){\circle*{50}}
 \put (400,200){\circle*{50}}
 \put (600,200){\circle*{50}}
 \put (100,374){\circle*{50}}
 \put (500,374){\circle*{50}}
 \put (  0,548){\circle*{50}}
 \put (200,548){\circle*{50}}
 \put (400,548){\circle*{50}}
 \put (600,548){\circle*{50}}
 \put (300,722){\circle*{50}}

 \color{black}
 \put (  0,548){\circle*{50}}
 \put (600,200){\circle*{50}}
 \put (300, 26){\circle*{50}}
 \put (300,722){\circle*{50}}
\end{picture}$
       & $\color{white}{{\Bigg|}}%
%
\setlength{\unitlength}{3947sp}
\begin{picture}(740,660)(-20,320)

 \color{magenta}
 \put (  0,174){\circle*{1}}
 \put (  5,184){\circle*{1}}
 \put ( 10,194){\circle*{1}}
 \put ( 15,204){\circle*{1}}
 \put ( 20,214){\circle*{1}}
 \put ( 25,224){\circle*{1}}
 \put ( 30,234){\circle*{1}}
 \put ( 35,244){\circle*{1}}
 \put ( 40,254){\circle*{1}}
 \put ( 45,264){\circle*{1}}
 \put ( 50,274){\circle*{1}}
 \put ( 55,284){\circle*{1}}
 \put ( 60,294){\circle*{1}}
 \put ( 65,304){\circle*{1}}
 \put ( 70,314){\circle*{1}}
 \put ( 75,324){\circle*{1}}
 \put ( 80,334){\circle*{1}}
 \put ( 85,344){\circle*{1}}
 \put ( 90,354){\circle*{1}}
 \put ( 95,364){\circle*{1}}

 \put (100,374){\circle*{1}}
 \put (105,384){\circle*{1}}
 \put (110,394){\circle*{1}}
 \put (115,404){\circle*{1}}
 \put (120,414){\circle*{1}}
 \put (125,424){\circle*{1}}
 \put (130,434){\circle*{1}}
 \put (135,444){\circle*{1}}
 \put (140,454){\circle*{1}}
 \put (145,464){\circle*{1}}
 \put (150,474){\circle*{1}}
 \put (155,484){\circle*{1}}
 \put (160,494){\circle*{1}}
 \put (165,504){\circle*{1}}
 \put (170,514){\circle*{1}}
 \put (175,524){\circle*{1}}
 \put (180,534){\circle*{1}}
 \put (185,544){\circle*{1}}
 \put (190,554){\circle*{1}}
 \put (195,564){\circle*{1}}

 \put (500,374){\circle*{1}}
 \put (495,384){\circle*{1}}
 \put (490,394){\circle*{1}}
 \put (485,404){\circle*{1}}
 \put (480,414){\circle*{1}}
 \put (475,424){\circle*{1}}
 \put (470,434){\circle*{1}}
 \put (465,444){\circle*{1}}
 \put (460,454){\circle*{1}}
 \put (455,464){\circle*{1}}
 \put (450,474){\circle*{1}}
 \put (445,484){\circle*{1}}
 \put (440,494){\circle*{1}}
 \put (435,504){\circle*{1}}
 \put (430,514){\circle*{1}}
 \put (425,524){\circle*{1}}
 \put (420,534){\circle*{1}}
 \put (415,544){\circle*{1}}
 \put (410,554){\circle*{1}}
 \put (405,564){\circle*{1}}


 \put (220,548){\circle*{10}}
 \put (230,548){\circle*{10}}
 \put (240,548){\circle*{10}}
 \put (250,548){\circle*{10}}
 \put (260,548){\circle*{10}}
 \put (270,548){\circle*{10}}
 \put (280,548){\circle*{10}}
 \put (290,548){\circle*{10}}
 \put (300,548){\circle*{10}}
 \put (310,548){\circle*{10}}
 \put (320,548){\circle*{10}}
 \put (330,548){\circle*{10}}
 \put (340,548){\circle*{10}}
 \put (350,548){\circle*{10}}
 \put (360,548){\circle*{10}}
 \put (370,548){\circle*{10}}
 \put (380,548){\circle*{10}}
 \put (390,548){\circle*{10}}
 \put (400,548){\circle*{10}}


 \put ( 20,200){\circle*{10}}
 \put ( 30,200){\circle*{10}}
 \put ( 40,200){\circle*{10}}
 \put ( 50,200){\circle*{10}}
 \put ( 60,200){\circle*{10}}
 \put ( 70,200){\circle*{10}}
 \put ( 80,200){\circle*{10}}
 \put ( 90,200){\circle*{10}}
 \put (100,200){\circle*{10}}
 \put (110,200){\circle*{10}}
 \put (120,200){\circle*{10}}
 \put (130,200){\circle*{10}}
 \put (140,200){\circle*{10}}
 \put (150,200){\circle*{10}}
 \put (160,200){\circle*{10}}
 \put (170,200){\circle*{10}}
 \put (180,200){\circle*{10}}
 \put (190,200){\circle*{10}}
 \put (200,200){\circle*{10}} 

 \put (220,200){\circle*{10}}
 \put (230,200){\circle*{10}}
 \put (240,200){\circle*{10}}
 \put (250,200){\circle*{10}}
 \put (260,200){\circle*{10}}
 \put (270,200){\circle*{10}}
 \put (280,200){\circle*{10}}
 \put (290,200){\circle*{10}}
 \put (300,200){\circle*{10}}
 \put (310,200){\circle*{10}}
 \put (320,200){\circle*{10}}
 \put (330,200){\circle*{10}}
 \put (340,200){\circle*{10}}
 \put (350,200){\circle*{10}}
 \put (360,200){\circle*{10}}
 \put (370,200){\circle*{10}}
 \put (380,200){\circle*{10}}
 \put (390,200){\circle*{10}}
 \put (400,200){\circle*{10}} 

 \put (420,200){\circle*{10}}
 \put (430,200){\circle*{10}}
 \put (440,200){\circle*{10}}
 \put (450,200){\circle*{10}}
 \put (460,200){\circle*{10}}
 \put (470,200){\circle*{10}}
 \put (480,200){\circle*{10}}
 \put (490,200){\circle*{10}}
 \put (500,200){\circle*{10}}
 \put (510,200){\circle*{10}}
 \put (520,200){\circle*{10}}
 \put (530,200){\circle*{10}}
 \put (540,200){\circle*{10}}
 \put (550,200){\circle*{10}}
 \put (560,200){\circle*{10}}
 \put (570,200){\circle*{10}}
 \put (580,200){\circle*{10}}
 \put (590,200){\circle*{10}}
 \put (600,200){\circle*{10}}
 
 \color{magenta}

 \put (610,200){\circle*{10}}
 \put (600,210){\circle*{10}}
 \put (590,220){\circle*{10}}
 \put (580,230){\circle*{10}}
 \put (570,240){\circle*{10}}
 \put (565,250){\circle*{10}}
 \put (560,260){\circle*{10}}
 \put (555,270){\circle*{10}}
 \put (550,280){\circle*{10}}
 \put (545,290){\circle*{10}}
 \put (540,300){\circle*{10}}
 \put (535,310){\circle*{10}}
 \put (530,320){\circle*{10}}
 \put (525,330){\circle*{10}}
 \put (520,340){\circle*{10}}
 \put (515,350){\circle*{10}}
 \put (510,360){\circle*{10}}
 \put (505,370){\circle*{10}}
 \put (500,374){\circle*{10}}
 \put (495,380){\circle*{10}} 

 \color{magenta}

 \color{magenta}
 \put (300,26){\circle*{50}}
 \put (  0,200){\circle*{50}}
 \put (200,200){\circle*{50}}
 \put (400,200){\circle*{50}}
 \put (600,200){\circle*{50}}
 \put (100,374){\circle*{50}}
 \put (500,374){\circle*{50}}
 \put (  0,548){\circle*{50}}
 \put (200,548){\circle*{50}}
 \put (400,548){\circle*{50}}
 \put (600,548){\circle*{50}}
 \put (300,722){\circle*{50}}

 \color{black}
 \put (  0,548){\circle*{50}}
 \put (600,548){\circle*{50}}
 \put (300, 26){\circle*{50}}
 \put (300,722){\circle*{50}}
\end{picture}$ & $\color{white}{{\Bigg|}}%
%
\setlength{\unitlength}{3947sp}
\begin{picture}(740,660)(-20,320)

 \color{magenta}
 \put (  0,174){\circle*{1}}
 \put (  5,184){\circle*{1}}
 \put ( 10,194){\circle*{1}}
 \put ( 15,204){\circle*{1}}
 \put ( 20,214){\circle*{1}}
 \put ( 25,224){\circle*{1}}
 \put ( 30,234){\circle*{1}}
 \put ( 35,244){\circle*{1}}
 \put ( 40,254){\circle*{1}}
 \put ( 45,264){\circle*{1}}
 \put ( 50,274){\circle*{1}}
 \put ( 55,284){\circle*{1}}
 \put ( 60,294){\circle*{1}}
 \put ( 65,304){\circle*{1}}
 \put ( 70,314){\circle*{1}}
 \put ( 75,324){\circle*{1}}
 \put ( 80,334){\circle*{1}}
 \put ( 85,344){\circle*{1}}
 \put ( 90,354){\circle*{1}}
 \put ( 95,364){\circle*{1}}



 \put (500,374){\circle*{1}}
 \put (505,384){\circle*{1}}
 \put (510,394){\circle*{1}}
 \put (515,404){\circle*{1}}
 \put (520,414){\circle*{1}}
 \put (525,424){\circle*{1}}
 \put (530,434){\circle*{1}}
 \put (535,444){\circle*{1}}
 \put (540,454){\circle*{1}}
 \put (545,464){\circle*{1}}
 \put (550,474){\circle*{1}}
 \put (555,484){\circle*{1}}
 \put (560,494){\circle*{1}}
 \put (565,504){\circle*{1}}
 \put (570,514){\circle*{1}}
 \put (575,524){\circle*{1}}
 \put (580,534){\circle*{1}}
 \put (585,544){\circle*{1}}
 \put (590,554){\circle*{1}}
 \put (595,564){\circle*{1}}
 \put (100,374){\circle*{1}}
 \put ( 95,384){\circle*{1}}
 \put ( 90,394){\circle*{1}}
 \put ( 85,404){\circle*{1}}
 \put ( 80,414){\circle*{1}}
 \put ( 75,424){\circle*{1}}
 \put ( 70,434){\circle*{1}}
 \put ( 65,444){\circle*{1}}
 \put ( 60,454){\circle*{1}}
 \put ( 55,464){\circle*{1}}
 \put ( 50,474){\circle*{1}}
 \put ( 45,484){\circle*{1}}
 \put ( 40,494){\circle*{1}}
 \put ( 35,504){\circle*{1}}
 \put ( 30,514){\circle*{1}}
 \put ( 25,524){\circle*{1}}
 \put ( 20,534){\circle*{1}}
 \put ( 15,544){\circle*{1}}
 \put ( 10,554){\circle*{1}}
 \put (  5,564){\circle*{1}} 

 \put (420,548){\circle*{10}}
 \put (430,548){\circle*{10}}
 \put (440,548){\circle*{10}}
 \put (450,548){\circle*{10}}
 \put (460,548){\circle*{10}}
 \put (470,548){\circle*{10}}
 \put (480,548){\circle*{10}}
 \put (490,548){\circle*{10}}
 \put (500,548){\circle*{10}}
 \put (510,548){\circle*{10}}
 \put (520,548){\circle*{10}}
 \put (530,548){\circle*{10}}
 \put (540,548){\circle*{10}}
 \put (550,548){\circle*{10}}
 \put (560,548){\circle*{10}}
 \put (570,548){\circle*{10}}
 \put (580,548){\circle*{10}}
 \put (590,548){\circle*{10}}
 \put (500,548){\circle*{10}} 

 \put (220,548){\circle*{10}}
 \put (230,548){\circle*{10}}
 \put (240,548){\circle*{10}}
 \put (250,548){\circle*{10}}
 \put (260,548){\circle*{10}}
 \put (270,548){\circle*{10}}
 \put (280,548){\circle*{10}}
 \put (290,548){\circle*{10}}
 \put (300,548){\circle*{10}}
 \put (310,548){\circle*{10}}
 \put (320,548){\circle*{10}}
 \put (330,548){\circle*{10}}
 \put (340,548){\circle*{10}}
 \put (350,548){\circle*{10}}
 \put (360,548){\circle*{10}}
 \put (370,548){\circle*{10}}
 \put (380,548){\circle*{10}}
 \put (390,548){\circle*{10}}
 \put (400,548){\circle*{10}}

 \put ( 20,548){\circle*{10}}
 \put ( 30,548){\circle*{10}}
 \put ( 40,548){\circle*{10}}
 \put ( 50,548){\circle*{10}}
 \put ( 60,548){\circle*{10}}
 \put ( 70,548){\circle*{10}}
 \put ( 80,548){\circle*{10}}
 \put ( 90,548){\circle*{10}}
 \put (100,548){\circle*{10}}
 \put (110,548){\circle*{10}}
 \put (120,548){\circle*{10}}
 \put (130,548){\circle*{10}}
 \put (140,548){\circle*{10}}
 \put (150,548){\circle*{10}}
 \put (160,548){\circle*{10}}
 \put (170,548){\circle*{10}}
 \put (180,548){\circle*{10}}
 \put (190,548){\circle*{10}}
 \put (200,548){\circle*{10}} 

 \put ( 20,200){\circle*{10}}
 \put ( 30,200){\circle*{10}}
 \put ( 40,200){\circle*{10}}
 \put ( 50,200){\circle*{10}}
 \put ( 60,200){\circle*{10}}
 \put ( 70,200){\circle*{10}}
 \put ( 80,200){\circle*{10}}
 \put ( 90,200){\circle*{10}}
 \put (100,200){\circle*{10}}
 \put (110,200){\circle*{10}}
 \put (120,200){\circle*{10}}
 \put (130,200){\circle*{10}}
 \put (140,200){\circle*{10}}
 \put (150,200){\circle*{10}}
 \put (160,200){\circle*{10}}
 \put (170,200){\circle*{10}}
 \put (180,200){\circle*{10}}
 \put (190,200){\circle*{10}}
 \put (200,200){\circle*{10}} 

 \put (220,200){\circle*{10}}
 \put (230,200){\circle*{10}}
 \put (240,200){\circle*{10}}
 \put (250,200){\circle*{10}}
 \put (260,200){\circle*{10}}
 \put (270,200){\circle*{10}}
 \put (280,200){\circle*{10}}
 \put (290,200){\circle*{10}}
 \put (300,200){\circle*{10}}
 \put (310,200){\circle*{10}}
 \put (320,200){\circle*{10}}
 \put (330,200){\circle*{10}}
 \put (340,200){\circle*{10}}
 \put (350,200){\circle*{10}}
 \put (360,200){\circle*{10}}
 \put (370,200){\circle*{10}}
 \put (380,200){\circle*{10}}
 \put (390,200){\circle*{10}}
 \put (400,200){\circle*{10}} 

 \put (420,200){\circle*{10}}
 \put (430,200){\circle*{10}}
 \put (440,200){\circle*{10}}
 \put (450,200){\circle*{10}}
 \put (460,200){\circle*{10}}
 \put (470,200){\circle*{10}}
 \put (480,200){\circle*{10}}
 \put (490,200){\circle*{10}}
 \put (500,200){\circle*{10}}
 \put (510,200){\circle*{10}}
 \put (520,200){\circle*{10}}
 \put (530,200){\circle*{10}}
 \put (540,200){\circle*{10}}
 \put (550,200){\circle*{10}}
 \put (560,200){\circle*{10}}
 \put (570,200){\circle*{10}}
 \put (580,200){\circle*{10}}
 \put (590,200){\circle*{10}}
 \put (600,200){\circle*{10}}
 
 \color{magenta}

 \put (610,200){\circle*{10}}
 \put (600,210){\circle*{10}}
 \put (590,220){\circle*{10}}
 \put (580,230){\circle*{10}}
 \put (570,240){\circle*{10}}
 \put (565,250){\circle*{10}}
 \put (560,260){\circle*{10}}
 \put (555,270){\circle*{10}}
 \put (550,280){\circle*{10}}
 \put (545,290){\circle*{10}}
 \put (540,300){\circle*{10}}
 \put (535,310){\circle*{10}}
 \put (530,320){\circle*{10}}
 \put (525,330){\circle*{10}}
 \put (520,340){\circle*{10}}
 \put (515,350){\circle*{10}}
 \put (510,360){\circle*{10}}
 \put (505,370){\circle*{10}}
 \put (500,374){\circle*{10}}
 \put (495,380){\circle*{10}} 

 \color{magenta}

 \color{magenta}
 \put (300,26){\circle*{50}}
 \put (  0,200){\circle*{50}}
 \put (200,200){\circle*{50}}
 \put (400,200){\circle*{50}}
 \put (600,200){\circle*{50}}
 \put (100,374){\circle*{50}}
 \put (500,374){\circle*{50}}
 \put (  0,548){\circle*{50}}
 \put (200,548){\circle*{50}}
 \put (400,548){\circle*{50}}
 \put (600,548){\circle*{50}}
 \put (300,722){\circle*{50}}

 \color{black}
 \put (300, 26){\circle*{50}}
 \put (300,722){\circle*{50}}
\end{picture}$ \\
       &  &  & $\color{white} {{.}}$                                                                \\
       &  &  & $\color{white} {{.}}$                                                                \\ \hline \hline
       $\color{white}{{\Bigg|}}%
%
\setlength{\unitlength}{3947sp}
\begin{picture}(740,660)(-20,320)

 \color{magenta}

 \put (100,374){\circle*{1}}
 \put (105,384){\circle*{1}}
 \put (110,394){\circle*{1}}
 \put (115,404){\circle*{1}}
 \put (120,414){\circle*{1}}
 \put (125,424){\circle*{1}}
 \put (130,434){\circle*{1}}
 \put (135,444){\circle*{1}}
 \put (140,454){\circle*{1}}
 \put (145,464){\circle*{1}}
 \put (150,474){\circle*{1}}
 \put (155,484){\circle*{1}}
 \put (160,494){\circle*{1}}
 \put (165,504){\circle*{1}}
 \put (170,514){\circle*{1}}
 \put (175,524){\circle*{1}}
 \put (180,534){\circle*{1}}
 \put (185,544){\circle*{1}}
 \put (190,554){\circle*{1}}
 \put (195,564){\circle*{1}}


 \put (500,374){\circle*{1}}
 \put (505,384){\circle*{1}}
 \put (510,394){\circle*{1}}
 \put (515,404){\circle*{1}}
 \put (520,414){\circle*{1}}
 \put (525,424){\circle*{1}}
 \put (530,434){\circle*{1}}
 \put (535,444){\circle*{1}}
 \put (540,454){\circle*{1}}
 \put (545,464){\circle*{1}}
 \put (550,474){\circle*{1}}
 \put (555,484){\circle*{1}}
 \put (560,494){\circle*{1}}
 \put (565,504){\circle*{1}}
 \put (570,514){\circle*{1}}
 \put (575,524){\circle*{1}}
 \put (580,534){\circle*{1}}
 \put (585,544){\circle*{1}}
 \put (590,554){\circle*{1}}
 \put (595,564){\circle*{1}}

 \put (420,548){\circle*{10}}
 \put (430,548){\circle*{10}}
 \put (440,548){\circle*{10}}
 \put (450,548){\circle*{10}}
 \put (460,548){\circle*{10}}
 \put (470,548){\circle*{10}}
 \put (480,548){\circle*{10}}
 \put (490,548){\circle*{10}}
 \put (500,548){\circle*{10}}
 \put (510,548){\circle*{10}}
 \put (520,548){\circle*{10}}
 \put (530,548){\circle*{10}}
 \put (540,548){\circle*{10}}
 \put (550,548){\circle*{10}}
 \put (560,548){\circle*{10}}
 \put (570,548){\circle*{10}}
 \put (580,548){\circle*{10}}
 \put (590,548){\circle*{10}}
 \put (500,548){\circle*{10}} 

 \put (220,548){\circle*{10}}
 \put (230,548){\circle*{10}}
 \put (240,548){\circle*{10}}
 \put (250,548){\circle*{10}}
 \put (260,548){\circle*{10}}
 \put (270,548){\circle*{10}}
 \put (280,548){\circle*{10}}
 \put (290,548){\circle*{10}}
 \put (300,548){\circle*{10}}
 \put (310,548){\circle*{10}}
 \put (320,548){\circle*{10}}
 \put (330,548){\circle*{10}}
 \put (340,548){\circle*{10}}
 \put (350,548){\circle*{10}}
 \put (360,548){\circle*{10}}
 \put (370,548){\circle*{10}}
 \put (380,548){\circle*{10}}
 \put (390,548){\circle*{10}}
 \put (400,548){\circle*{10}}



 \put (220,200){\circle*{10}}
 \put (230,200){\circle*{10}}
 \put (240,200){\circle*{10}}
 \put (250,200){\circle*{10}}
 \put (260,200){\circle*{10}}
 \put (270,200){\circle*{10}}
 \put (280,200){\circle*{10}}
 \put (290,200){\circle*{10}}
 \put (300,200){\circle*{10}}
 \put (310,200){\circle*{10}}
 \put (320,200){\circle*{10}}
 \put (330,200){\circle*{10}}
 \put (340,200){\circle*{10}}
 \put (350,200){\circle*{10}}
 \put (360,200){\circle*{10}}
 \put (370,200){\circle*{10}}
 \put (380,200){\circle*{10}}
 \put (390,200){\circle*{10}}
 \put (400,200){\circle*{10}} 

 \put (420,200){\circle*{10}}
 \put (430,200){\circle*{10}}
 \put (440,200){\circle*{10}}
 \put (450,200){\circle*{10}}
 \put (460,200){\circle*{10}}
 \put (470,200){\circle*{10}}
 \put (480,200){\circle*{10}}
 \put (490,200){\circle*{10}}
 \put (500,200){\circle*{10}}
 \put (510,200){\circle*{10}}
 \put (520,200){\circle*{10}}
 \put (530,200){\circle*{10}}
 \put (540,200){\circle*{10}}
 \put (550,200){\circle*{10}}
 \put (560,200){\circle*{10}}
 \put (570,200){\circle*{10}}
 \put (580,200){\circle*{10}}
 \put (590,200){\circle*{10}}
 \put (600,200){\circle*{10}}
 
 \color{magenta}
 \put (210,200){\circle*{10}}
 \put (200,210){\circle*{10}}
 \put (190,220){\circle*{10}}
 \put (180,230){\circle*{10}}
 \put (170,240){\circle*{10}}
 \put (165,250){\circle*{10}}
 \put (160,260){\circle*{10}}
 \put (155,270){\circle*{10}}
 \put (150,280){\circle*{10}}
 \put (145,290){\circle*{10}}
 \put (140,300){\circle*{10}}
 \put (135,310){\circle*{10}}
 \put (130,320){\circle*{10}}
 \put (125,330){\circle*{10}}
 \put (120,340){\circle*{10}}
 \put (115,350){\circle*{10}}
 \put (110,360){\circle*{10}}
 \put (105,370){\circle*{10}}
 \put (100,374){\circle*{10}}
 \put ( 95,380){\circle*{10}}

 \put (610,200){\circle*{10}}
 \put (600,210){\circle*{10}}
 \put (590,220){\circle*{10}}
 \put (580,230){\circle*{10}}
 \put (570,240){\circle*{10}}
 \put (565,250){\circle*{10}}
 \put (560,260){\circle*{10}}
 \put (555,270){\circle*{10}}
 \put (550,280){\circle*{10}}
 \put (545,290){\circle*{10}}
 \put (540,300){\circle*{10}}
 \put (535,310){\circle*{10}}
 \put (530,320){\circle*{10}}
 \put (525,330){\circle*{10}}
 \put (520,340){\circle*{10}}
 \put (515,350){\circle*{10}}
 \put (510,360){\circle*{10}}
 \put (505,370){\circle*{10}}
 \put (500,374){\circle*{10}}
 \put (495,380){\circle*{10}} 

 \color{magenta}

 \color{magenta}
 \put (300,26){\circle*{50}}
 \put (  0,200){\circle*{50}}
 \put (200,200){\circle*{50}}
 \put (400,200){\circle*{50}}
 \put (600,200){\circle*{50}}
 \put (100,374){\circle*{50}}
 \put (500,374){\circle*{50}}
 \put (  0,548){\circle*{50}}
 \put (200,548){\circle*{50}}
 \put (400,548){\circle*{50}}
 \put (600,548){\circle*{50}}
 \put (300,722){\circle*{50}}

 \color{black}
 \put (  0,548){\circle*{50}}
 \put (  0,200){\circle*{50}}
 \put (300, 26){\circle*{50}}
 \put (300,722){\circle*{50}}
\end{picture}$ & $\color{white}{{\Bigg|}}\input{mmkxa14-36_julia3f.latex}$
       & $\color{white}{{\Bigg|}}\input{mmkxa14-36_julia3g.latex}$ & $\color{white}{{\Bigg|}}\input{mmkxa14-36_julia3h.latex}$ \\
       &  &  & $\color{white} {{.}}$                                                                \\
       &  &  & $\color{white} {{.}}$                                                                \\ \hline \hline
\end{tabular}
  \caption{
           A list of all eight possible type loops surrounding the central hexagon 
           within a star-shaped cell, up to rotational symmetry. 
           Including all possible rotations of each type, there are 32 distinct loops. 
           Each dimerization realizes exactly one of these 32 loops, 
           where the links of the loop alternate between occupied and unoccupied, 
           yielding two possible realizations via dimers for each loop. 
           The hexagons shown in Fig. \ref{figure_666}(a) are surrounded by loops of the 
           types  shown in the last column. 
}
\label{table3}
\end{table}

The special choice 
$t_{1} = \ldots = t_{32} = V_{1} = \ldots = V_{32} > 0$ 
is an instance of a Rokhsar-Kivelson (RK) point. 
Here, the ground state is the equal amplitude superposition of all admissable dimer coverings 
\begin{eqnarray}\label{state_psi}
        |\Psi\rangle = \sum _{D}|D\rangle\,,
\end{eqnarray}
where, for PBCs the sum may be restricted to one topological sector, thus leading to a four-fold ground state degeneracy. 
On the kagome lattice, this RK-point lies in the interior of a $\mathbb{Z}_{2}$ topological 
phase \cite{misguich,seidel2009linear,PRLwildeboer} and is fully integrable \cite{misguich}, 
owing to the fact that the sums of the operators 
in \eqref{ham1}  associated to any given star will commute for different stars. 
Furthermore, for the kagome lattice, the entanglement entropy of the states \eqref{state_psi} 
can be analytically calculated and  shown to display area law entanglement entropy \cite{wildeboer}. 

{\it The scar kagome dimer model. ---} 
The goal of this Letter is to design a system made of dimer degrees of freedom 
on the kagome lattice 
that admits quantum many-body scar states in its spectrum. 
We begin by observing that the states \eqref{state_psi} are annihilated by the Hamiltonian 
\eqref{ham1} not only at the special integrable point $t_i = V_i = 1$, but whenever $t_i = V_i$. 
This is so because each local term associated with $t_i = V_i$ annihilates \Eq{state_psi}. 
Moving away from the integrable point while preserving $t_i = V_i$ destroys the integrability  
(all eigenstates {\em except} \Eq{state_psi} will not be known analytically), but preserves 
the fact that \Eq{state_psi} is an exact zero energy mode. For positive $t_i = V_i$, all associated 
local operators thus have a {\em common} ground state in \Eq{state_psi}. This is then also the ground state of $H$, 
the latter being the sum of these local operators. It is then common to call $H$ a {\em frustration free} Hamiltonian. 
\begin{figure}
\includegraphics[width=1.00\columnwidth]{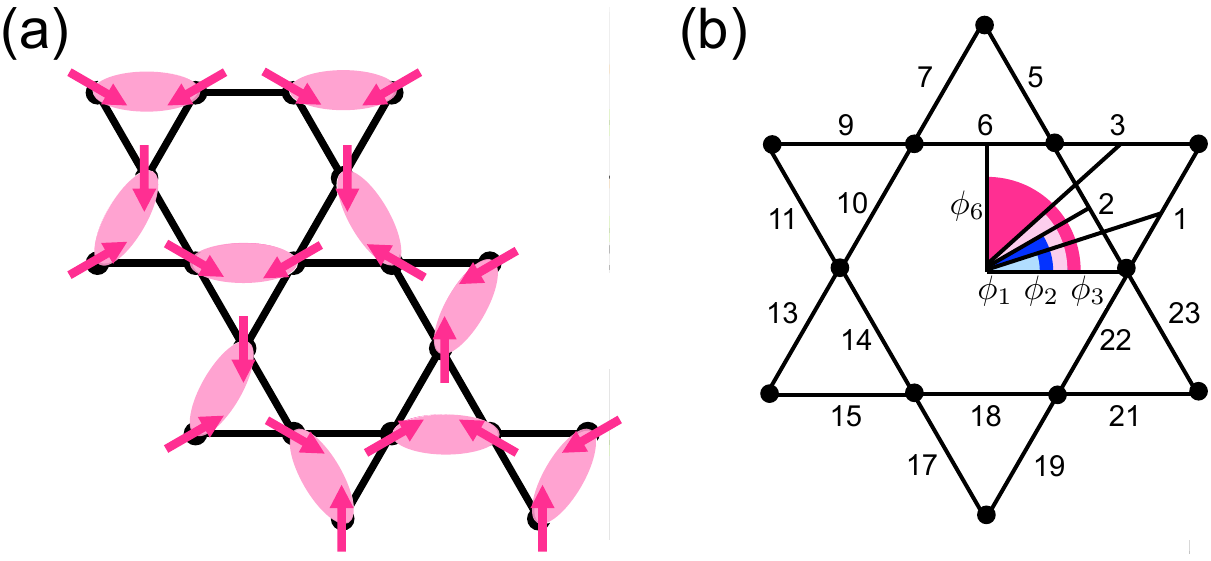}
\caption{
         (a) A possible dimer covering and the analogous arrow representation
             first introduced by Zeng and Elser \cite{zeng1995quantum}: The number of 
             incoming arrows at each triangle must be even 
             (0 or 2), and dimers are associated to links between two incoming arrows.
         (b) Labeling convention for the links of the 12-site star-shaped cells of the kagome lattice.
             The 12-site kagome star consists of a total of 18 links. 
             We label each link by a number $l$ such that its angle bisector 
             from the midpoint of the hexagon makes the angle 
             $\phi_{l} = \pi l/12$ ($l$ ``skips'' multiples of $4$) with the x-axis, as shown.  
}
\label{figure_666}
\end{figure}

The following strategy is expected to work generally for frustration free Hamiltonians (though not always while preserving all 
symmetries): We introduce $t_i = V_i\equiv \alpha_i$, and choose the $\alpha_i$ different and {\em not all of the same sign}. 
\Eq{state_psi} is still a zero mode of the resulting Hamiltonian, but it is not a ground state, but rather a state somewhere 
in the middle of the spectrum. 
We establish that this state is a true quantum many-body scar by observing the following properties. 
First, the state itself satisfies area-law entanglement, despite being highly excited. 
This is usually inferred from the fact that it is the ground state of {\em some} local Hamiltonian, 
and it is analytically provable for the kagome lattice state \eqref{state_psi} considered here \cite{wildeboer} . 
We further show numerically  that the surrounding states in the energy spectrum behave ``generically'', 
i.e., have much larger entanglement entropy (expected to be volume law in the thermodynamic limit), and  
satisfy the expected level statistics appropriate to the respective symmetry sector they lie in. 
This in particular means that the Hamiltonian is not ``special'' in the sense of integrability.
 
Explicitly, we introduce a scar dimer model Hamiltonian as follows 
\begin{eqnarray}\label{ham2}
{\cal H}^{scar} =  
\sum_{\textrm{\ding{65}}}
\sum_{\ell=1}^{32} \alpha_{\ell}
\left( \ket{D_{\ell}} - \ket{\overline{D_{\ell}}} \right)
\left( \bra{D_{\ell}} - \bra{\overline{D_{\ell}}} \right)
\;. 
\end{eqnarray}
The sums in \eqref{ham2} go over all 12-site kagome stars and over all $32$ loop coverings. 
$D_{\ell}$ and $\overline{D_{\ell}}$ represent the dimerizations associated with loop $\ell$. 
We could easily follow the strategy described above while preserving all lattice 
symmetries. However, the level statistics we are interested in make sense only within symmetry sectors, 
as there is no level repulsion rule between different sectors. To avoid an over-abundance of symmetry sectors, 
we preserve only translational symmetry by choosing 
\begin{eqnarray}\label{alphal} 
\alpha_{\ell} = C + \sum_{l \in \rm loop}{\rm sin}\left(5 \cdot \phi_{l} + \delta\right)\;.
\end{eqnarray}
which simulates the influence of a substrate with 5-fold rotational symmetry. Here,
$l$ refers to the link labels defined in the caption of Fig. \ref{figure_666} along
with the associated angles $\phi_{l}$. 
We choose $C = -0.05$ to make dimer-loops with inversion 
symmetry contribute, and $\delta = 0.1$ to render the mirror axes of the ``substrate'' 
different from those of the lattice.
\begin{figure}
        \includegraphics[width=1.00\columnwidth]{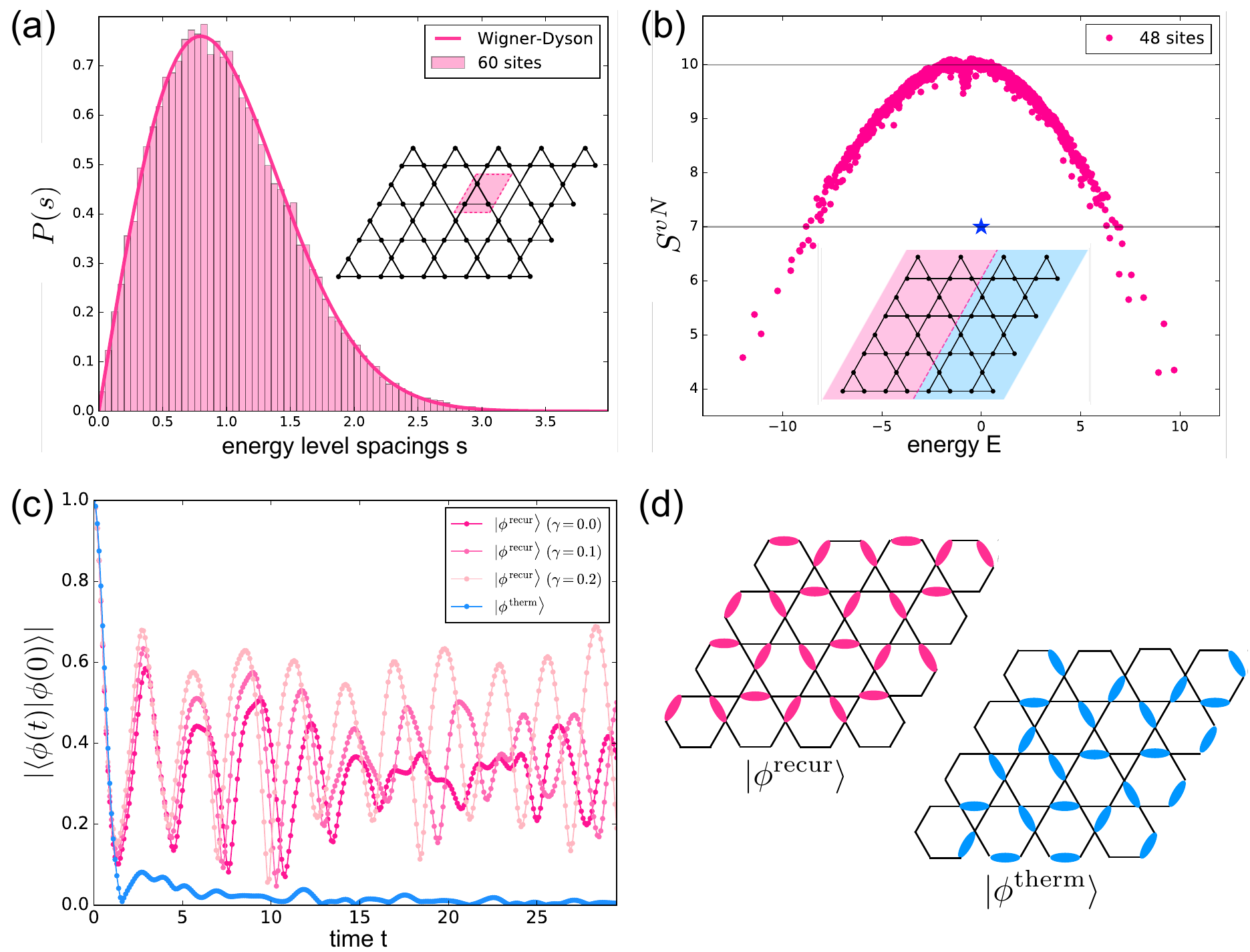} 
\caption{
         (a) Distribution of energy levels for the time-reversal invariant zero-momentum,
         $(W_x,W_y) = (0,0)$ topological sector, which contains a scar state \eqref{state_psi}.
         The inset shows the 60-site kagome lattice (with unit cell shaded) used in the calculation.
         An unfolding technique using 4378 groups containing 12 energies each has been used
         for binning the data (cf., e.g., Ref. \cite{matsui}).
         The resulting data closely resemble a GOE distribution (solid curve),
         indicating that almost all states thermalize.
         (b) The von Neumann entanglement entropy for 
         all states within the zero momentum sector of topological winding numbers $(0,0)$ 
         for a 48 site kagome lattice, bi-partitioned into two 24-site ``ribbons'' (inset). 
         The scar state has $S^{\sf vN} = 7$ and is marked by a blue star. 
         Thermalizing eigenstates of similar energy are well separated and have $S^{\sf vN} \approx 10$. 
         (c) Overlap ${\cal O}(t)= |\langle \phi (0) | \phi(t)\rangle|$ for a 
         special initial configuration $|\phi^{\rm recur}\rangle$  (see (d)), 
         both for the original Hamiltonian and deformations parametrized by $\gamma$ (see text). 
         Pronounced recurrence phenomena are observed, clearly distinct from a typical 
         initial configuration ($|\phi^{\rm therm}\rangle$, shown in blue), where the overlap decays rapidly. 
         (d) The special, non-thermalizing initial configuration  $|\phi^{\rm recur}\rangle$  
         along with the typical, thermalizing state $|\phi^{\rm therm}\rangle$ discussed in (c). 
         Other non-thermalizing initial configuration are related to the one shown here by lattice symmetries.  
}
\label{figure_levels}
\end{figure}

{\it Level statistics. ---}
While the scar state of the model and its properties are analytically under control, we proceed by 
numerically investigating the genericity of its other levels. We focus on (translational) symmetry sectors 
with time reversal symmetry, which contain the scar state. 
Fig \ref{figure_levels}(a) shows the distribution 
of energy eigenvalues for a 60 site kagome lattice with PBCs, 
within the zero-momentum sector that has the scar state located roughly in the middle of the spectrum (see Fig \ref{figure_levels}(b)). 
Here, we work within the $(W_x, W_y) = (0,0)$ topological sector and use an unfolding technique to bin the data (cf., e.g., \cite{matsui}).  
One observes that the distribution is well described by the Gaussian orthogonal ensemble (GOE), 
as expected for generic real matrices. This can be quantified further as follows \cite{OganesyanHuse, atas2013distribution}. 
Introducing level spacings $s_n=E_n-E_{n-1}$, one defines quantities $r_n= s_n/s_{n-1}$ and $\tilde r_n=\min(r_n, 1/r_n)$. 
With this, we find the average of the quantities $\tilde r_n$ over 
all 
symmetry-inequivalent 
time-reversal invariant  
momentum sectors within the $(W_x, W_y) = (0,0)$ topological sector to be 
$\langle \tilde r \rangle = 0.5333$. 
This is quite close  to the exact value of $\tilde r_{\textrm{GOE}} = 0.5359$ \cite{rigol}, 
and markedly different from the corresponding value 
$\tilde r_{\textrm{Poisson}} = 0.3863$ \cite{OganesyanHuse} for the Poisson distribution. 
The average $r_n$ value tends to require larger samples owing to the possibility of small denominators, 
but at $\langle  r \rangle =  1.7626$ is likewise very close to the exact value of 
$r_{\textrm{GOE}} = 1.7781$. 
Had we at least retained inversion symmetry, all symmetry sectors would be described by real matrices, 
and one would expect to find similar values in all sectors. 
However, inversion being absent, there are time-reversal non-invariant momentum sectors in this model, 
not containing the scar state \eqref{state_psi}, which, for sufficiently generic models, 
can be expected to be described by the Gaussian unitary ensemble. 
To test this, we carried out the analogous analysis for these sectors, 
finding $\tilde r = 0.5996$ and $r = 1.3709$, again very close to the exact values 
$\tilde r_{\textrm{GUE}} = 0.60266$ and $r_{\textrm{GUE}} = 1.3607$. 
These findings lend strong support to the hypothesis that the majority of the high energy states in the  
spectrum of Hamiltonian \eqref{ham2} are ergodic, i.e., they thermalize.  

{\it Entanglement entropy. ---}
To complement the above findings, we calculate bipartite entanglement entropy 
for all states of the scar-containing symmetry sector (fixing also the topological sector) 
for a 48 site kagome lattice with PBCs. By their definition, quantum many-body scar states belong to the 
bulk of the spectrum while simultaneously  violating the ETH, i.e., they fail to thermalize and 
display low (sub-volume) entanglement behavior. 
In contrast, generic high-energy states do thermalize and exhibit a volume-dependent  
entanglement behavior. 
We find that this contrast is starkly displayed already on the 48 site lattice, which we cut 
into two 24 site ribbons wrapping around the torus (Fig. \ref{figure_levels}(b), inset). 
For simplicity, in doing so we regard the arrows of the Zeng-Elser representation of 
permissible dimerizations of the kagome lattice as the physical local degrees of freedom (Fig. \ref{figure_666}(b)). 
For the ribbon described, whose boundary passes eight unit cells on each side, and in the 
presence of the topological sector constraint, one may show that the (base 2) von Neumann entanglement entropy, 
$S^{\sf vN} = -\sum_w  w \log_2 w$, where the sum goes over the eigenvalues $w$ 
of the reduced density matrix, equals $7$. Here, the sum goes over the eigenvalues of the local 
density matrix of the ribbon. 
Fig. \ref{figure_levels}(b) clearly shows that the scar state (blue star) is isolated from the rest 
of the spectrum (purple dots) in terms of its much lower entanglement as compared to surrounding bulk energy eigenstates. 
This establishes the state \eqref{state_psi} as a bona fide quantum many-body scar. 

{\it Additional scars and fidelity dynamics. ---} 
Multiple features in Fig.~\ref{figure_levels}(b) suggest the presence of additional scars 
in the spectrum that, 
while less removed from the continuum of eigenstates than \Eq{state_psi}, are nonetheless 
distinct in their entanglement properties. 
We test this hypothesis by studying fidelity dynamics. 
Fig.~\ref{figure_levels}(c) shows the overlap ${\cal O}(t)= |\langle \phi (0) | \phi(t)\rangle|$ 
for different initial states $|\phi(0)\rangle$ and their time evolved counterparts, $|\phi(t)\rangle$. 
All initial states are chosen to be simple product states of dimers. 
While for most such initial configurations, the overlap ${\cal O}(t)$ rapidly decays 
to zero (the ``thermalizing'' state $|\phi^{\rm therm}\rangle$ in Fig.\ref{figure_levels}(d)), some special initial states show remarkable recurrence oscillations. 
This is, in particular, true for the initial configuration $|\phi^{\rm recur}\rangle$ shown in Fig.\ref{figure_levels}(d)
and configurations related to that by lattice symmetries (even though the latter are not symmetries
of ${\cal H}^{scar}$). 
It is clear that the scar state \eqref{state_psi} cannot by itself be responsible for the observed
oscillations of $|\phi^{\rm recur}\rangle$: Being the equal amplitude superposition of all dimer basis states, 
it does not render any particular basis states special. Moreover, the oscillations 
displayed by the special initial states require multiple isolated eigenstates at different energies 
to have exceptional overlaps with these initial states. The observed 
phenomenology is indeed highly consistent with the presence of multiple 
scars \cite{turner2018quantum,turner2018weak,iadecola2019quantum,schecteria,lin2020quantum,wildeboer2021a,wildeboer2021b}. Moreover, we find that it is stable toward 
small perturbations that remove the existence of an RK-eigenstate:
To this end, we increase the $V_i$-terms (as defined in \Eq{ham1}) of \Eq{ham2} by a fraction $\gamma$
relative to the $t_i$-terms:
As shown in  Fig. \ref{figure_levels}(c), the recurrence phenomena and associated oscillations
easily survive an increase by 20\%.

{\it Conclusion. ---} 
We investigated a general approach to turning classes of frustration free 
lattice Hamiltonians into ones containing isolated quantum many-body scars 
in their spectrum while retaining most or all symmetries. 
In addition, the introduction of disorder is straightforward, as is the generalization to other lattices. 
We applied this strategy to a two-dimensional quantum dimer model on the kagome lattice, 
retaining full translational symmetry. 
We demonstrated that this model contains an exactly known quantum many-body scar with 
analytically accessible entanglement properties. 
We established that the remainder of the eigenstates and energy spectrum exhibit no ``fine-tuned'' behavior. 
Specifically, for a 60-site kagome lattice, we showed that bulk energies conform to the Gaussian ensembles 
expected for their respective symmetry sectors, and we 
calculated von Neumann entanglement entropies for all states within the scar-sector 
of a 48 site kagome lattice, exposing the scar's isolated character. 
Due to their quality of being numerically manageable on 
fairly large-size lattices, quantum dimer models of the type considered here 
should become a fertile playground for investigations of this kind. 
Indeed, the existence of revival phenomena stable
toward generic perturbations suggests that the
large parameter space of the class of models given here
will prove fruitful for further studies of equilibration processes
in 2D lattice models.
Lastly,  the original, frustration-free quantum dimer model 
stabilizes a $\mathbb{Z}_2$ topological phase. 
This lends a topological character to our scar states.
It will be of interest to contrast this with the universality class of the ground
state of our model, which is currently unknown. 
We are hopeful that these observations will stipulate further investigation. 
\begin{acknowledgments}
AS is indebted to Z. Nussinov for insightful discussions.
Work by A.S. has been supported by the National Science Foundation Grant No. DMR-2029401.
O.E.  acknowledges  support  from  the  National  Science Foundation Grant No. DMR-1904716.
\end{acknowledgments}
\bibliography{bibfile1}
\end{document}